\documentclass[preprint]{aastex61}
\usepackage{amsmath,amstext}
\usepackage[T1]{fontenc}
\usepackage{apjfonts} 
\usepackage[figure,figure*]{hypcap}

\def\deg{$^{\rm o}$}

\def\arcsec{\ifmmode '' \else $''$\fi}

\def\arcsecpoint{\ifmmode ''\!. \else $''\!.$\fi}

\def\kms{\ifmmode {\rm km\ s}^{-1} \else km s$^{-1}$\fi}
\def\Msun{\ifmmode {\rm M}_{\odot} \else M$_{\odot}$\fi}
\def\Lsun{\ifmmode {\rm L}_{\odot} \else L$_{\odot}$\fi}
\def\Zsun{\ifmmode {\rm Z}_{\odot} \else Z$_{\odot}$\fi}

\def\ergscm2{ergs\,s$^{-1}$\,cm$^{-2}$}
\def\icm3{{\rm cm}^{-3}}
\def\icm2{{\rm cm}^{-2}}
\def\qo{\ifmmode q_{\rm o} \else $q_{\rm o}$\fi}
\def\Ho{\ifmmode H_{\rm o} \else $H_{\rm o}$\fi}
\def\ho{\ifmmode h_{\rm o} \else $h_{\rm o}$\fi}

\def\vFWHM{\ifmmode v_{\mbox{\tiny FWHM}} \else
            $v_{\mbox{\tiny FWHM}}$\fi}

\def\gtorder{\mathrel{\raise.3ex\hbox{$>$}\mkern-14mu
             \lower0.6ex\hbox{$\sim$}}}
\def\ltorder{\mathrel{\raise.3ex\hbox{$<$}\mkern-14mu
             \lower0.6ex\hbox{$\sim$}}}
\def\proptwid{\mathrel{\raise.3ex\hbox{$\propto$}\mkern-14mu
             \lower0.6ex\hbox{$\sim$}}}

\newcommand{\Ly}{Ly$\alpha$}
\newcommand{\Ha}{H$\alpha$}
\newcommand{\Hb}{H$\beta$}
\newcommand{\Hg}{H$\gamma$}

\shorttitle{DEIMOS Survey in COSMOS}
\shortauthors{Hasinger, G. et al.}

\begin{document}

\title{The DEIMOS 10k spectroscopic survey catalog of the COSMOS field\footnote{The data presented herein were obtained at the W. M. Keck Observatory, which is operated as a scientific partnership among the California Institute of Technology, the University of California and the National Aeronautics and Space Administration. The Observatory was made possible by the generous financial support of the W. M. Keck Foundation.}}

\author{Hasinger, G.}
\affiliation{University of Hawaii, Institute for Astronomy, 2680 Woodlawn Dr., Honolulu, HI, 96822, USA}
\affiliation{European Space Astronomy Centre (ESA/ESAC), Director of Science, E-28691 Villanueva de la Ca\~nada, Madrid, Spain}
\author{Capak, P.}
\affiliation{IPAC, California Institute of Technology, 1200 E. California\ Blvd., Pasadena, CA, 91125, USA}
\author{Salvato, M.}
\affiliation{Max Planck Institut f\"ur extraterrestrische Physik, Giessenbachstrasse 1, D-85748, Garching bei M\"unchen, Germany}
\affiliation{Excellence Cluster Universe, Boltzmannstrasse 1, 85748 Garching}
\author{Barger, A. J.}
\affiliation{Department of Astronomy, University of Wisconsin-Madison, 475 North Charter Street, Madison, WI 53706, USA}
\affiliation{University of Hawaii, Institute for Astronomy, 2680 Woodlawn Dr., Honolulu, HI, 96822, USA}
\author{Cowie, L. L.}
\affiliation{University of Hawaii, Institute for Astronomy, 2680 Woodlawn Dr., Honolulu, HI, 96822, USA}
\author{Faisst, A.}
\affiliation{IPAC, California Institute of Technology, 1200 E. California\ Blvd., Pasadena, CA, 91125, USA}
\author{Hemmati, S.}
\affiliation{IPAC, California Institute of Technology, 1200 E. California\ Blvd., Pasadena, CA, 91125, USA}
\author{Kakazu, Y.}
\affiliation{Subaru Telescope, 650 N. A'ohoku Place, Hilo, HI 96720, USA}
\author{Kartaltepe, J.}
\affiliation{School of Physics and Astronomy, Rochester Institute of Technology, 54 Lomb Memorial Drive, Rochester, NY 14623, USA}
\author{Masters, D.}
\affiliation{IPAC, California Institute of Technology, 1200 E. California\ Blvd., Pasadena, CA, 91125, USA}
\author{Mobasher, B.}
\affiliation{University of California, Riverside, 900 University Ave, Riverside, CA 92697, USA}
\author{Nayyeri, H.}
\affiliation{Department of Physics and Astronomy, University of California Irvine, Irvine 92521, CA, USA}
\author{Sanders, D.}
\affiliation{University of Hawaii, Institute for Astronomy, 2680 Woodlawn Dr., Honolulu, HI, 96822, USA}
\author{Scoville, N.~Z.}
\affiliation{California Institute of Technology, 1200 E. California\ Blvd., Pasadena, CA, 91125, USA}
\author{Suh, H.}
\affiliation{University of Hawaii, Institute for Astronomy, 2680 Woodlawn Dr., Honolulu, HI, 96822, USA}
\affiliation{Subaru Telescope, 650 N. A'ohoku Place, Hilo, HI 96720, USA}
\author{Steinhardt, C.}
\affiliation{DAWN Center, Niels Bohr Institute, Blegdamsvej 17, DK-2100 K\o benhavn, Denmark}
\affiliation{Dark Cosmology Centre, Niels Bohr Institute, Blegdamsvej 17, DK-2100 K\o benhavn, Denmark}
\author{Yang, Fengwei}
\affiliation{Department of Physics and Laboratory for Space Research, The University of Hong Kong, Hong Kong SAR, China}

\begin{abstract}
We present a catalog of 10718 objects in the COSMOS field observed through multi-slit spectroscopy with the Deep Imaging Multi-Object Spectrograph (DEIMOS) on the Keck II telescope in the wavelength range $\sim$ 5500-9800 \AA. The catalog contains 6617 objects with high-quality spectra (two or more spectral features), and 1798 objects with a single spectroscopic feature confirmed by the photometric redshift. For 2024 typically faint objects we could not obtain reliable redshifts. The objects have been selected from a variety of input catalogs based on multi-wavelength observations in the field, and thus have a diverse selection function, which enables the study of the diversity in the galaxy population. The magnitude distribution of our objects is peaked at $I_{AB}\sim23$ and $K_{AB}\sim21$, with a secondary peak at $K_{AB}\sim24$. We sample a broad redshift distribution in the range $0<z<6$, with one peak at $z\sim1$, and another one around $z\sim4$.  We have identified 13 redshift spikes at $z>0.65$ with chance probabilities $<4\times10^{-4}$, some of which are clearly related to protocluster structures of sizes $>10 \ {\rm Mpc}$. An object-to-object comparison with a multitude of other spectroscopic samples in the same field shows that our DEIMOS sample is among the best in terms of fraction of spectroscopic failures and relative redshift accuracy. We have determined the fraction of spectroscopic blends to about 0.8\% in our sample. This is likely a lower limit and at any rate well below the most pessimistic expectations. Interestingly, we find evidence for strong lensing of \Ly \ background emitters within the slits of 12 of our target galaxies, increasing their apparent density by about a factor of 4.

\end{abstract}


\section{Introduction}
The Cosmic Evolution Survey \citep[COSMOS;][]{2007ApJS..172....1S} is a galaxy survey designed to probe the formation and evolution  of galaxies, star formation, and active galactic nuclei (AGNs), both over cosmic time (redshift $z=0.5-6$) and as a function of the local galaxy environment defined by the dark matter and its large-scale structure (LSS). It is designed to be representative of the large scale structure of the universe defined by the dark matter scaffolding. The survey covers a 2 deg$^2$ equatorial field with multiwavelength imaging and spectroscopy from X-ray to radio wavelengths by most of the major space-based telescopes (Hubble, Spitzer, GALEX, XMM, Chandra, Herschel, NuStar) and large ground based observatories (Keck, Subaru, VLA, ESO-VLT, UKIRT, NOAO, CFHT, JCMT, ALMA and others). Over 2 million galaxies are detected in deep optical images \citep{2009ApJ...690.1236I}, and 1.2 million in the NIR \citep{2016ApJS..224...24L}, spanning 75\% of the age of the Universe. A subset of the field has also been selected as part of the ``Cosmic Assembly Near-infrared Deep Extragalactic Legacy Survey'' (CANDELS), and surveyed deeper in the NIR with Hubble \citep{2017ApJS..228....7N}. Given the depth and resolution of these data, COSMOS also provides unprecedented samples of rare objects and structures at high redshifts with greatly reduced cosmic variance, compared to other, smaller area deep surveys.  A comprehensive description of the large-scale structure in the COSMOS Field has been presented by \cite{2007ApJS..172..150S,2013ApJS..206....3S}, and \cite{2007Natur.445..286M}.

\vskip 0.1 truecm
Spectroscopic identifications are an essential part of every large-scale cosmological survey, since redshifts and distances are required for practically all physical applications of the data. Massive spectroscopy campaigns have been undertaken in the COSMOS field over the last decade \citep[e.g.,][]{2007ApJS..172...70L,2009ApJ...696.1195T,2009ApJS..184..218L,2011ApJ...741....8C,2015A&A...576A..79L,2015ApJS..220...12S}. Here we describe the more than 100 multi-object spectroscopy observations of a total of more than 10,000 objects taken with the medium-resolution Deep Imaging Multi-Object Spectrograph \citep[DEIMOS;][]{2003SPIE.4841.1657F} on the Keck II telescope during the decade 2006-2016. 
Although the DEIMOS redshifts have already been used for a wide variety of fundamental astrophysical studies, this paper is the first comprehensive presentation of the entire spectrosopic data. 

\vskip 0.1 truecm
DEIMOS spectra have e.g. been used to calibrate the excellent photometric redshifts for galaxies \citep{2009ApJ...690.1236I,2013A&A...556A..55I,2016ApJS..224...24L} and AGN  \citep{2009ApJ...690.1250S,2011ApJ...742...61S,2016ApJ...817...34M} in the COSMOS field. They have also been used to characterize optical counterparts of Spitzer MIPS \citep{2010ApJ...709..572K,2010ApJ...721...98K} and Herschel PACS sources \citep{2013ApJ...778..131L}, as well as X-ray sources discovered by XMM-Newton \citep[e.g.][]{2009ApJ...693....8B,2010ApJ...716..348B,2012ApJ...759....6E} and Chandra \citep[e.g.][]{2012ApJS..201...30C,2016ApJ...817...34M,2016ApJ...827..150M} and to determine the most accurate cosmic evolution of AGN luminosity functions so far \citep[e.g.][]{2012ApJ...755..169M,2015ApJ...804..104M}. The 3D spatial correlation function of galaxies and AGN, as well as their Halo occupation function can provide strong constraints on the type of dark matter halos, in which the objects typically reside \citep{2009A&A...494...33G,2011ApJ...736...99A,2012ApJ...758...47A,2014ApJ...796....4A,2016ApJ...832...70A}. Optical spectroscopy has been used to determine the black hole masses in broad emission line objects and thus to measure the Eddington ratios of their accretion flows \citep{2015ApJ...815..129S}. The spectra have been used to identify high-redshift \Ly \ emitters \citep{2012ApJ...760..128M} and high-redshift protoclusters in the COSMOS field \citep{2011Natur.470..233C}, allowing a study of the metal properties of high-redshift galaxies \citep[e.g.][]{2016ApJ...822...29F}, as well as designing high redshift ALMA surveys \citep[e.g.][]{2015Natur.522..455C}. Finally, the DEIMOS spectra will be used to calibrate future surveys with ESA's Euclid and the NASA Wide-Field Infrared Survey-Telescope (WFIRST) missions \citep{2017ApJ...841..111M} with far reaching cosmological implications.

\vskip 0.1 truecm
In this paper we present the overall observations in Section 2 and the spectral analysis in Section 3. The description of the released catalog is presented in Section 4 while the comparison of various metrics between different selection functions is discussed in Section 5. Examples of scientific applications of the catalog are presented in Section $6-7$ and the Summary in Section 8 concludes the paper. Throughout this work we adopt a $\Lambda$-cosmology with $\Omega_M=0.3$ and $\Omega_\Lambda=0.7$, and $H_0=70 km \ s^{−1} \ Mpc^{-1}$ \citep{2003ApJS..148..175S}, and all magnitudes are given in the AB system.

% TABLE 1:  Selected Samples
% \scriptsize
\begin{deluxetable*}{lrrrrrrrrr}
% \tablewidth{0pt}
\tablecaption{Subsamples of objects included in slit masks}
\tablehead{
\colhead{Sample} &
\colhead{Total} &
\colhead{Unique} &
\colhead{Q=2} &
\colhead{Q=1.5} &
\colhead{Q=1} &
\colhead{Q=0} &
\colhead{Success} &
\colhead{$<I_{AB}>$} &
\colhead{$<K_{AB}>$} }

\startdata
X-ray         & 1237 &  589 &  697 &  305 &  50 & 185 & 56.3\% & 23.0 & 21.0 \\
high-z        & 2007 & 1878 &  564 &  468 & 141 & 834 & 28.1\% & 25.2 & 23.7 \\
MIPS          & 3001 & 1588 & 2201 &  420 &  45 & 335 & 73.3\% & 22.0 & 20.4 \\
VLA           & 1238 &  136 &  827 &  231 &  27 & 171 & 66.8\% & 22.5 & 20.5 \\
Herschel      &  787 &  171 &  626 &  118 &   3 &  40 & 79.5\% & 22.2 & 20.8 \\
OVV           &  363 &  314 &  255 &   63 &   8 &  37 & 70.2\% & 23.5 & 22.2 \\
OII           &  262 &  238 &  184 &   45 &   1 &  32 & 70.2\% & 24.2 & 22.9 \\
PL AGN        &   69 &   29 &   38 &   18 &   2 &  11 & 55.1\% & 23.0 & 21.6 \\
Filler        & 3188 & 3188 & 2158 &  384 &  69 & 577 & 67.7\% & 22.0 & 20.9 \\
Serendipitous &  966 &  810 &  763 &  134 &  12 &  57 & 78.9\% & 22.3 & 21.3 \\
{\bf Total}   & {\bf 10718} & & {\bf 6617} & {\bf 1798} & {\bf 279} & {\bf 2924} & {\bf 61.7\%} & {\bf 22.9} & {\bf 21.4} \\
\enddata
\label{table:subsamples}
\tablecomments{Shown here is the total number of objects and the number of unique objects in each subsample and each quality class Q (see below). Many objects fall in more than one sub-sample as discussed in the text. Success is the percentage of high-quality spectra (Q=2) compared to the total in each subsample. The last two columns are the mean I- and K-magnitudes of the subsample.}
\end{deluxetable*}
% \normalsize

\section{Observations}

The spectroscopic observations in the COSMOS field were conducted with DEIMOS on the Keck II telescope. The Field of View (FOV) of DEIMOS is approximately $16 \times 4 \ {\rm arcmin}^2$ which allows placement of slit masks in the field for multi-object spectroscopy of $\sim 60-100$ objects. The slit masks for our observations were prepared with the IRAF ``dsimulator'' software provided by the Keck observatory\footnote{\url{http://www2.keck.hawaii.edu/inst/deimos/dsim.html}}. For most masks we used a minimum slit length of 10\arcsec \ with a gap between slits of 0.35\arcsec. The slit width was typically $1\arcsec$. The samples of objects placed on the slit masks were selected according to different criteria, depending on scientific objectives of the corresponding program. Table \ref{table:subsamples} gives a summary of the different subsamples we have grouped our targets in. 

\begin{itemize}

\item Spectroscopy of Spitzer mid-Infrared ($24\,{\rm \mu m}$ and $70\,{\rm \mu m}$) MIPS sources (PI: Kartaltepe). This contains a sample of ULIRGs selected through their mid-IR Spitzer/MIPS detections \citep[see][]{2010ApJ...709..572K}. 

\item Deep spectroscopy of high redshift candidates selected through a variety of broad-band and narrow-band photometry \citep[see also][]{2012ApJ...760..128M}; this subsample is denoted ``high-z'' (PI: Capak).
These objects are potential Lyman break galaxies (LBGs) and \Ly \
emitters (LAEs) at $z\sim4.2$, $z\sim4.8$, and $z\sim5.6$ selected from
intermediate and narrowband Subaru SuprimCam observations. 

\item A subset of narrow- and intermediate-band excess sources selected possible [OII] emission line objects in the redshift range $0.3<z<1.6$; this is called the ``OII'' subsample \citep{2007ApJS..172..456T}. The authors discriminated against other possible strong emission lines using the broad-band colors.

\item Spectroscopy of the optical counterparts of X-ray sources selected from XMM-Newton \citep{2007ApJS..172...29H} and Chandra \citep[e.g.,][]{2012ApJS..201...30C,2016ApJ...817...34M,2016ApJ...827..150M} surveys (PI: Hasinger); this is the ``X-ray'' subsample.

\item Additional slits were assigned to other sources with lower space density. These include radio sources (``VLA''), Herschel PACS sources (``Herschel'') and power-law AGN candidates \citep[PL AGN,][]{2017MNRAS.466L.103C,2012ApJ...748..142D}
.
\item One subsample of the spectroscopic targets consists of optically variable sources (``OVV''), as defined in \cite{2009ApJ...690.1250S}. In the course of the determination of photometric redshifts for AGN, the object photometry is corrected for variability. Using the same procedure, potentially variable objects have been selected here for follow-up spectroscopy.

\item Slits in empty mask areas were assigned to ``Filler'' targets; these were predominantly drawn from a z-band magnitude-limited sample weighted to photometric redshifts at $z>0.8$, where the zCOSMOS program \citep{2009ApJS..184..218L} had difficulties due to fringing. The zCOSMOS BRIGHT sample contains spectroscopy for a sample of 10,644
galaxies with I(AB) < 22.5 mag. Another sample of filler targets were subthreshold Chandra X-ray detections. 

\item Serendipitous sources were picked up on the target slits and form the ``Serendipitous'' subsample. 

\end{itemize}

\begin{figure}
\begin{center}
\includegraphics[angle=0,scale=.57]{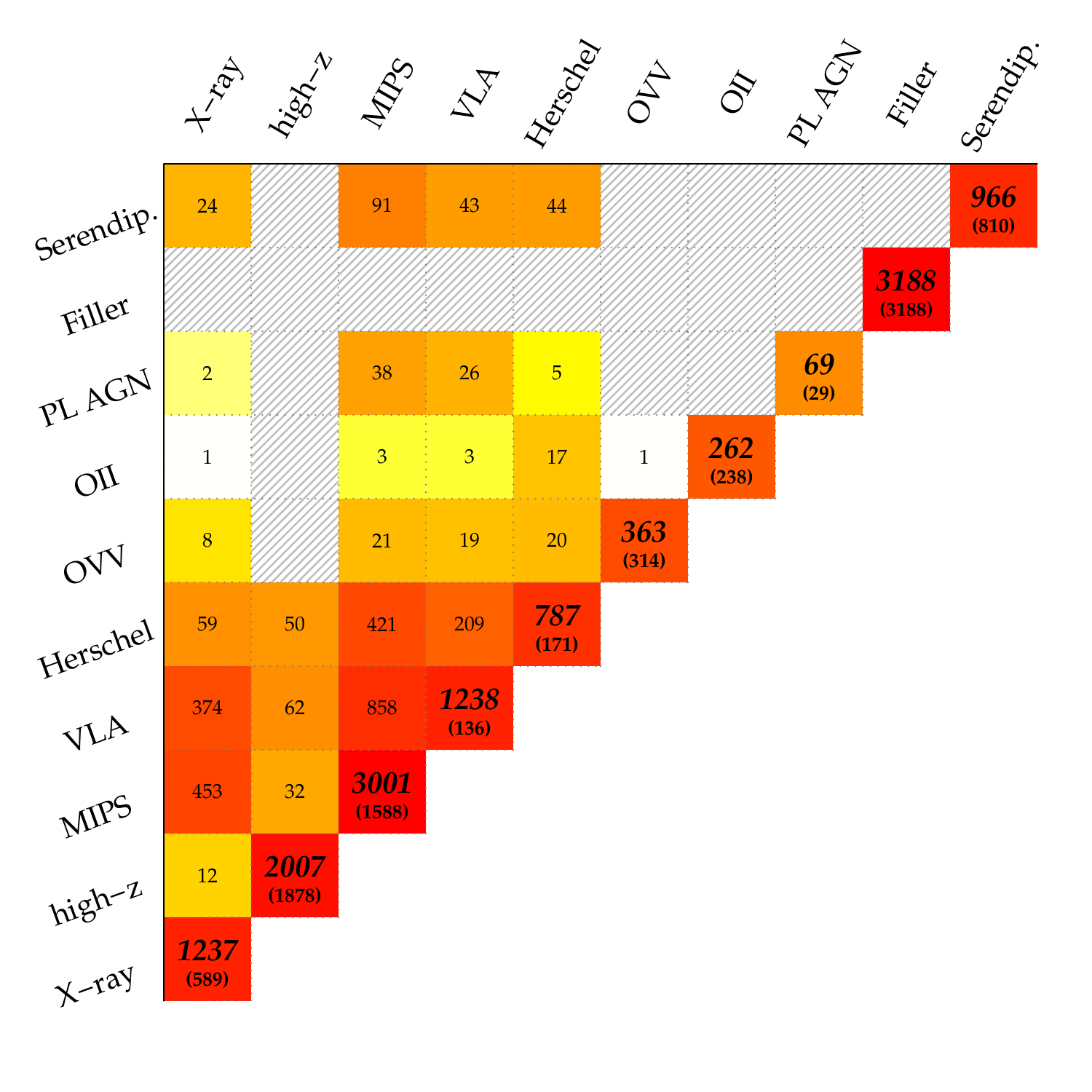}
\caption{Sample crosscorrelation matrix. \label{fig0}}
\end{center}
\end{figure}

This sample selection is not unique in each subsample, because a particular object can appear in several subsamples. There was also an evolution of the various subsamples over the years. When e.g. deeper X-ray or VLA observations became available across the whole COSMOS field, some objects were added to the respective subsamples. Figure \ref{fig0} shows the overlap matrix between the various subsamples. The diagonal shows the total number and number of unique objects in each subsample (also noted in table \ref{table:subsamples}), while the off-diagonal elements show the respective overlap with other subsamples. The entries are color coded with increasing sample sizes from white, to yellow, to orange to red.

\vskip 0.1 truecm
Compared to other, more homogeneous, sample selections for multi-object spectroscopy, e.g. the magnitude-selected samples observed with VIMOS on the VLT \citep{2009ApJS..184..218L,2015A&A...576A..79L}, our overall spectroscopic sample is much more heterogeneous. This, however, has the advantage that some of the magnitude-selection biases are washed out, and that our sample spans an unprecedented range in redshifts and magnitudes, as well as other characteristics (see below). 

The observing log, covering the years $2007-2017$, is shown in Table \ref{table:slitmasks}. The full list can be accessed online on the official COSMOS webpage\footnote{\url{http://cosmos.astro.caltech.edu}}. The typical seeing for these observations was $\sim0.7-1.2"$. For of the observations the 600ZD or 830G gratings were used, with blue blocking filters GG455 and OG550, respectively. The 600ZD grating yields a wavelength coverage of $\sim4800-10000$\AA \ with a dispersion of 0.65 \AA/pixel and a spectral resolution of $R\sim2000$. The 830G grating yields a wavelength coverage of $\sim 6700-10500$\AA \ with a dispersion of 0.47 \AA/pixel and a spectral resolution of $R\sim2700$. The wavelength coverage varies with the position of the respective slit on the mask. The spectral resolution is sufficient to e.g. distinguish the [OII] 3727\AA \ doublet emission line from a skewed \Ly \ profile and thus provides secure redshifts even in single emission line cases. Each mask was observed with a total integration time of $1-2$ hours, typically split into 4 exposures with an ABBA dither pattern of $\pm2"$. Depending on exposure times and spectral content we reach limiting magnitudes of $I_{AB}=23.5-25$. The last column in Table \ref{table:slitmasks} shows the number of slits assigned for each mask, as well as the number of successful high quality spectroscopic IDs, and serendipitous sources (see below).

% TABL% TABLE 1:  SLITMASKS
\scriptsize
%long table for AAStex
\startlongtable
%switch to this for emulateapj
%\LongTables
\begin{deluxetable*}{lcccccccccl}
% \tablewidth{0pt}
\tablecaption{List of observed slitmasks.}
\tablehead{
\colhead{Name} &
\colhead{R.A.} &
\colhead{Dec.} &
\colhead{PA} &
\colhead{Date} &
\colhead{UTC} &
\colhead{Exp.} &
\colhead{Airmass} &
\colhead{Grating} &
\colhead{Filter} &
\colhead{Spectra} \\
\colhead{} &
\colhead{(J2000)} &
\colhead{(J2000)} &
\colhead{(\deg)} &
\colhead{} &
\colhead{} &
\colhead{(hrs)} &
\colhead{} &
\colhead{} &
\colhead{} &
\colhead{} }

\startdata
clae1 & 9:59:49 & +02:07:43.7 & 6.5 & 1/14/07 & 9:09:53 & 2 & 1.79 & 830G & OG550 & 75/56/5 \\
clae2 & 10:01:36 & +02:31:52.9 & 66.1 & 1/14/07 & 15:11:22 & 2 & 1.28 & 830G & OG550 & 77/66/7 \\
clae3 & 10:00:53 & +02:35:14.7 & 82.9 & 1/15/07 & 10:31:03 & 2 & 1.25 & 830G & OG550 & 80/77/16 \\
clae4 & 10:02:24 & +01:55:20.5 & 74.7 & 1/15/07 & 14:19:21 & 2 & 1.14 & 830G & OG550 & 71/62/9 \\
clae5 & 10:00:41 & +01:33:42.1 & 94.6 & 1/16/07 & 10:55:26 & 2 & 1.17 & 830G & OG550 & 74/64/11 \\
clae6 & 9:59:48 & +01:38:16.2 & -15.9 & 1/16/07 & 15:49:07 & 1 & 1.54 & 830G & OG550 & 72/78/20 \\
clae7 & 10:01:35 & +01:45:44.9 & 10.7 & 1/17/07 & 11:19:07 & 2 & 1.12 & 830G & OG550 & 75/62/11 \\
clae8 & 10:00:58 & +01:52:32.1 & 60.1 & 1/17/07 & 12:27:01 & 2 & 1.05 & 830G & OG550 & 77/43/1 \\
clae9 & 10:00:05 & +01:54:13.0 & 20.7 & 1/18/07 & 8:02:42 & 2 & 2.69 & 830G & OG550 & 69/64/14 \\
clae10 & 10:03:05 & +01:55:25.7 & 0.1 & 1/18/07 & 15:02:28 & 2 & 1.31 & 830G & OG550 & 71/70/15 \\
B7 & 9:59:41 & +02:27:47.7 & 26.0 & 1/21/07 & 7:40:47 & 1 & 2.96 & 600ZD & GG455 & 117/58/0 \\
F7 & 9:59:41 & +02:27:48.0 & 26.0 & 1/21/07 & 9:51:39 & 1.2 & 1.31 & 600ZD & GG455 & 122/61/0 \\
... \\
m3be & 10:02:04 & +02:39:52.5 & 113.5 & 2/22/15 & 6:41:50 & 1 & 1.75 & 600ZD & GG455 & 62/48/4 \\
m40be & 10:00:37 & +01:37:11.4 & 113.5 & 2/22/15 & 7:59:35 & 1 & 1.27 & 600ZD & GG455 & 67/77/13 \\
m28be & 9:58:20 & +02:09:11.3 & 113.5 & 2/22/15 & 9:10:57 & 1 & 1.09 & 600ZD & GG455 & 60/50/5 \\
m4bd & 9:59:27 & +02:51:25.1 & 113.5 & 2/22/15 & 10:23:16 & 1 & 1.05 & 600ZD & GG455 & 75/76/12 \\
m37be & 10:01:24 & +01:36:44.6 & 113.5 & 2/22/15 & 10:49:58 & 1 & 1.06 & 600ZD & GG455 & 67/72/10 \\
m7be & 10:02:30 & +02:31:02.2 & 113.5 & 2/22/15 & 12:37:02 & 1 & 1.27 & 600ZD & GG455 & 68/71/10 \\
mn46 & 9:59:40 & +01:35:24.5 & 113.0 & 11/10/15 & 13:36:49 & 1 & 1.70 & 600ZD & GG455 & 69/74/12 \\
mn47 & 9:58:47 & +02:38:22.1 & 32.0 & 11/10/15 & 15:23:12 & 1 & 1.15 & 600ZD & GG455 & 68/63/6 \\
mn44 & 9:57:42 & +02:18:09.5 & 148.0 & 11/11/15 & 13:29:48 & 1 & 1.69 & 600ZD & GG455 & 68/66/6 \\
mn45 & 9:58:50 & +02:22:06.3 & 48.0 & 11/11/15 & 14:42:46 & 1 & 1.27 & 600ZD & GG455 & 67/72/9 \\
mn48 & 10:01:51 & +02:46:41.4 & 165.0 & 11/11/15 & 15:08:48 & 1 & 1.19 & 600ZD & GG455 & 71/60/6 \\
mn50 & 10:02:40 & +02:07:27.5 & 46.0 & 5/30/16 & 5:55:25 & 1 & 1.22 & 600ZD & GG455 & 72/62/8 \\
mn51 & 9:58:11  & +02:07:06.9 & 52.0 & 10/27/17 & 13:52:14 & 1.3 & 2.14 & 600ZD & GG455 & 74/57/6 \\
\enddata
\label{table:slitmasks}
\tablecomments{The last column shows the number of spectra in each mask. The first entry is the number of slits assigned, the second entry is the number of high-quality redshifts ($Q\ge1.5$) obtained, and the third entry is the number of serendipitous sources in each mask. The full list can be retrieved at \url{http://cosmos.astro.caltech.edu}.}
\end{deluxetable*}
\normalsize

\begin{figure}
\includegraphics[angle=0,scale=.55]{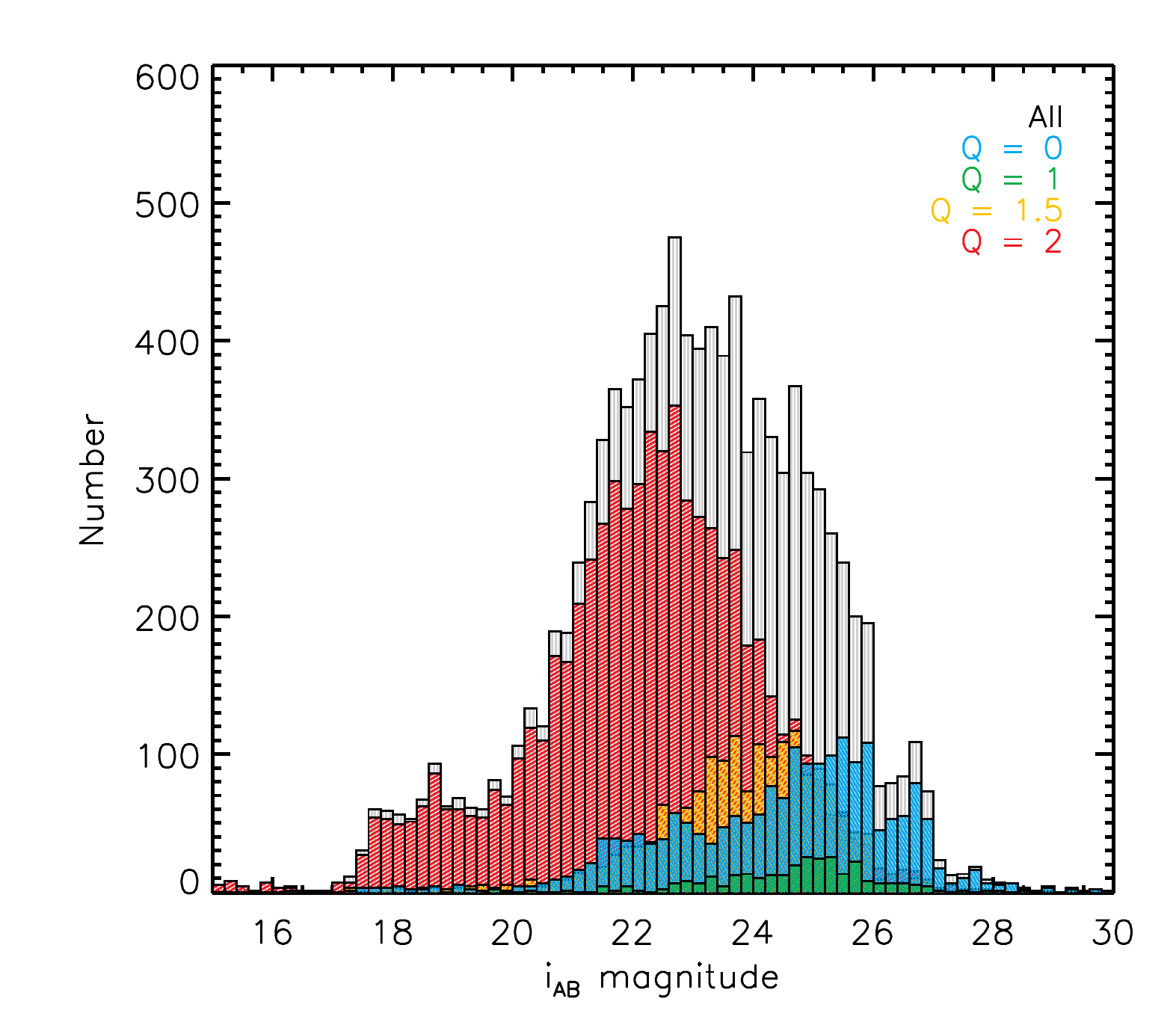}
\includegraphics[angle=0,scale=.55]{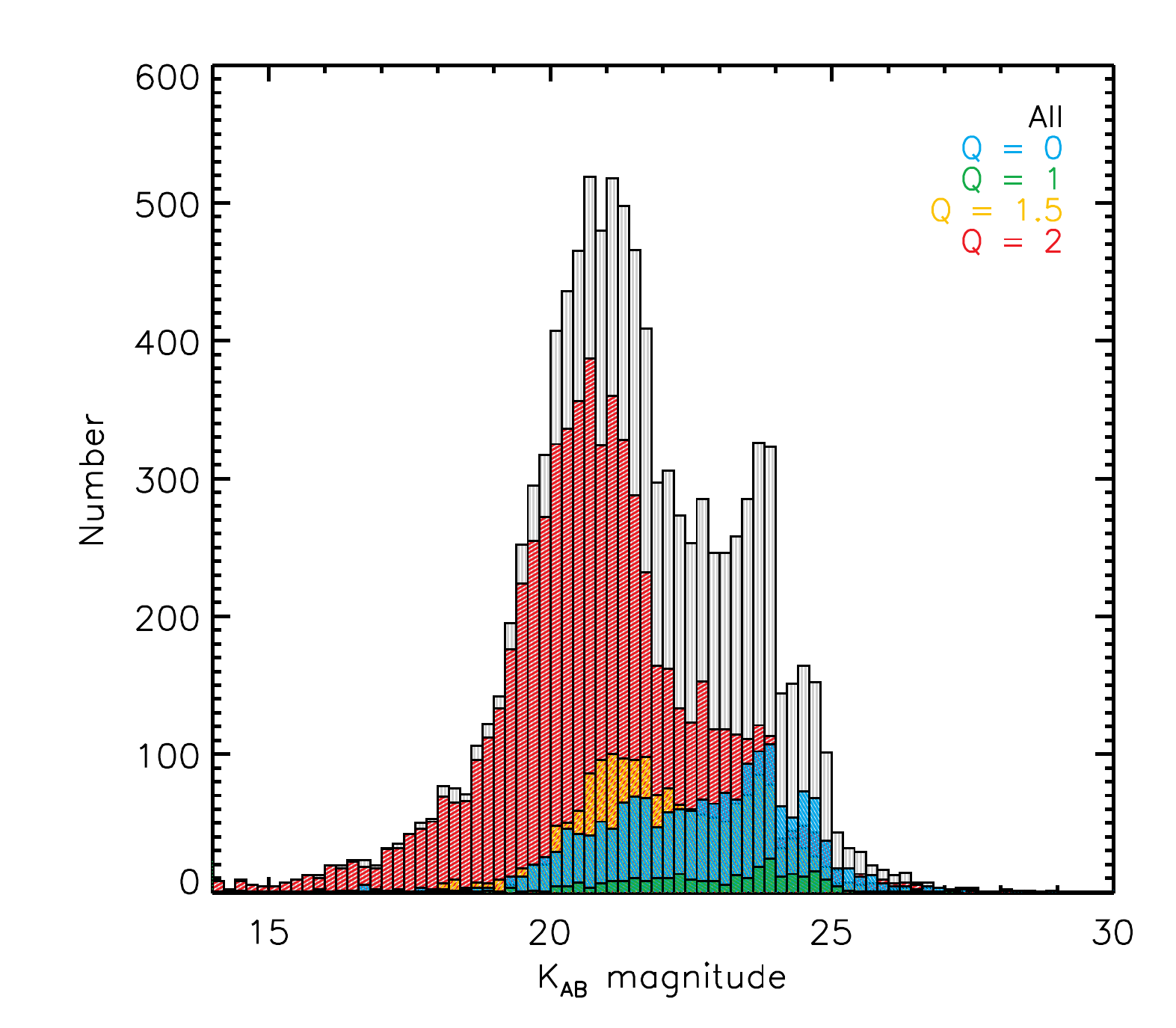}
\caption{Histogram of the magnitude distribution of objects with different spectral quality classes (see text). The highest quality spectra (red) cover brighter optical/NIR magnitudes. Sources for which we failed to assign a redshift (in blue) are faint in the observed magnitudes. The gray histogram shows all sources.\label{fig1}}
\end{figure}

\begin{figure}
\includegraphics[angle=0,scale=.55]{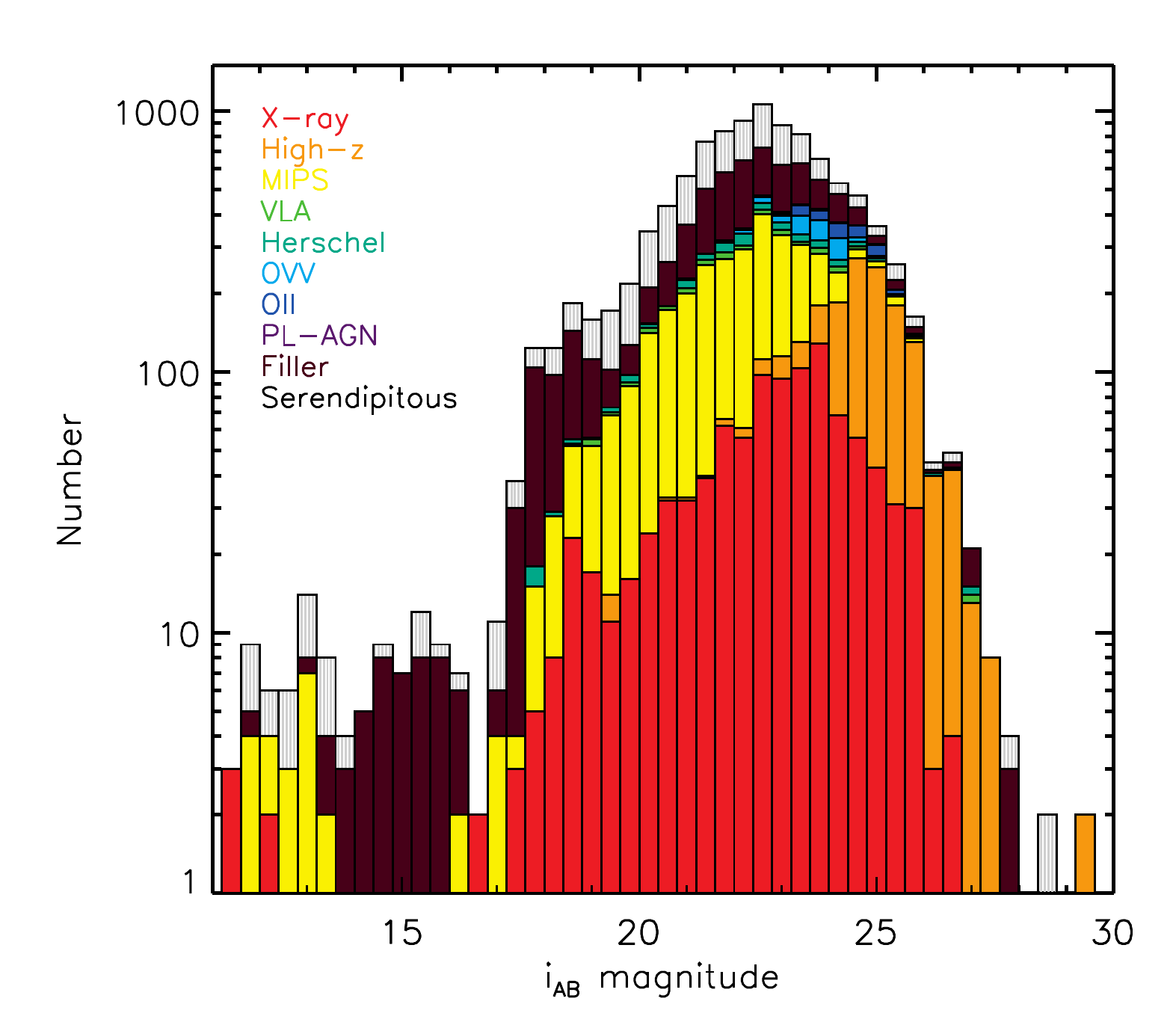}
\includegraphics[angle=0,scale=.55]{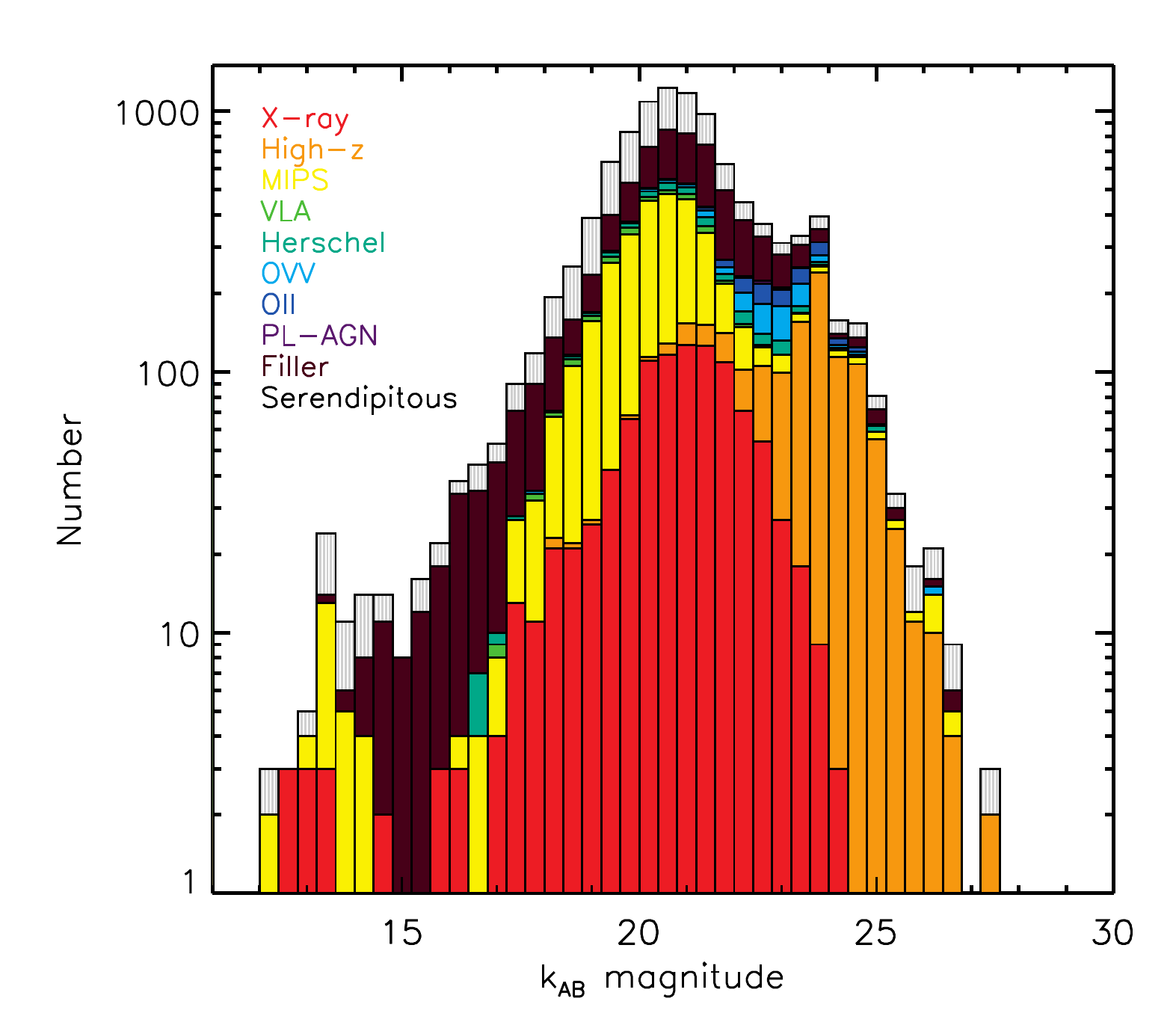}
\caption{Stacked histograms for $I_{AB}$ and $K_{AB}$ magnitude distributions for the different subsamples of sources with reliable redshift (Q=2). The high-redshift candidates (orange) show the faintest optical/NIR magnitudes.\label{fig2}}
\end{figure}

\section{Spectral analysis}

Most of the raw data were reduced using our specifically modified version of the DEEP2 data reduction pipeline. The original DEEP2 pipeline \citep[{\em{spec2d}};][]{2012MNRAS.419.3018C,2013ApJS..208....5N} consists of the bias removal, flat-fielding, slit-tilt correction, cosmic ray rejection, sky subtraction, and wavelength calibration. The modified version accounts for dithering, removes the ghosting on the grating data, and corrects for variable slit losses and errors in the alignment introduced by the dithering. The flux calibration was then applied by using the existing multi-wavelength photometry available on the COSMOS field.

\begin{figure}
\begin{center}
\includegraphics[angle=0,scale=.7]{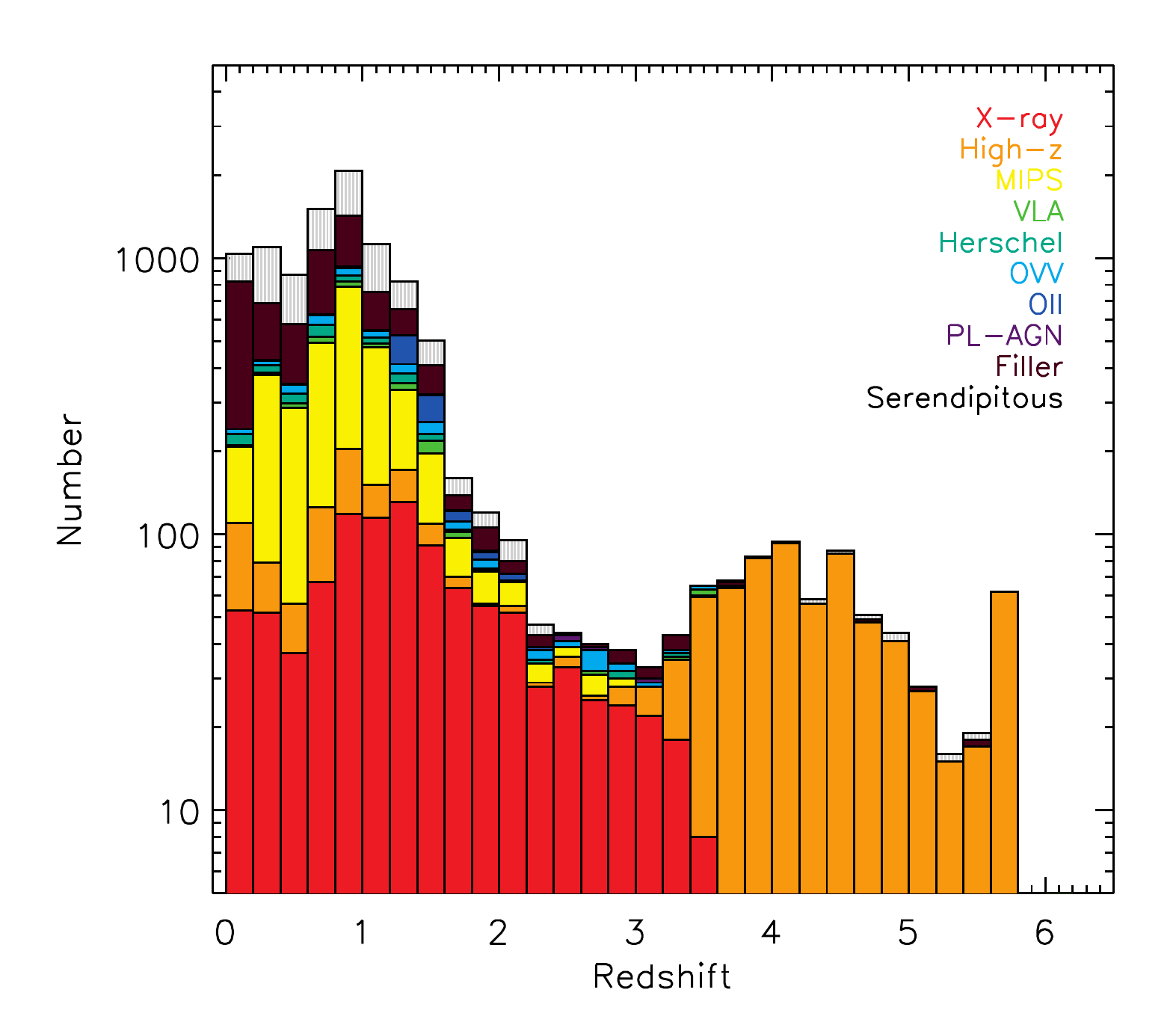}
\caption{Redshift distribution of the different subsamples of objects on the DEIMOS masks.\label{fig3}}
\end{center}
\end{figure}

\vskip 0.1 truecm
For most of the spectroscopic analysis and redshift identification we used the interactive IDL program ``{\em SpecPro}'' developed for viewing and analyzing astronomical spectra in the context of the COSMOS survey \citep{2011PASP..123..638M}. Its interactive design lets the user simultaneously view spectroscopic, photometric, and imaging data, allowing for rapid object classification and redshift determination. The spectroscopic redshift can be determined by automated cross-correlation of continuum and spectral features against a variety of spectral templates or by manually overlaying common emission and absorption features on the 1-D and 2-D spectra. Stamp images and the spectral energy distribution (SED) of a source can be displayed with the interface, with the positions of prominent photometric features indicated on the SED plot. {\em SpecPro} also displays the slit position on the 2D spectra and stamp images, and enables the re-extraction of 1D spectra from the 2D data. This is in particular important for serendipitous sources on the slits, which can be analyzed and positionally identified using {\em SpecPro}. Such serendipitous detections were later identified in the photometric master catalogs \citep{2009ApJ...690.1236I, 2016ApJS..224...24L}. For the quality assessment of the spectroscopic redshifts, we employed a scheme based on the zCOSMOS spectroscopic survey \citep{2009ApJS..184..218L}, where $Q_f=4$ is the highest quality spectrum with several identified spectral features, $Q_f=3$ corresponds to a lower quality, but still reliable spectroscopic identification with at least two spectral features or a single characteristically shaped emission line like a skewed \Ly \ or a double-humped [OII] line. $Q_f=2$ and $Q_f=1$ are lower quality spectra with decreasing reliability. $Q_f=9$ is based on a single high-significance emission line. If there are broad lines in the spectrum, the quality flag is increased by 10 (i.e. 14, 13, 12, 11, 19). If a source was detected serendipitously, the quality flag was increased by 20 (i.e. 24, 23, 22, 21, 29). Every spectrum was reviewed by at least two co-authors to find a consolidated solution in case of ambiguities. 

\vskip 0.1 truecm
For further discussion we define a more comprehensive quality flag ``Q'', which combines spectroscopic and photometric redshift information, following \cite{2004ApJS..155...73Z}. The $Q_f$ flags 3, 4, 13, 14, 23, 24 are given the value $Q=2$, signaling reliable spectroscopic identification. The $Q_f$ flags 1, 2, 9, 11, 12, 19, 21, 22, 29 are given the value $Q=1$ for an uncertain spectroscopic identification. However, if the photometric redshift value for a $Q=1$ source is matching with the uncertain spectroscopic redshift within an interval $\delta_z/(1+z)<0.1$, where $\delta_z=|z_{spec}-z_{phot}|$, we raise the quality flag to $Q=1.5$. An unsuccessful redshift measurement yields $Q=0$. Figure \ref{fig1} shows the magnitude distribution for samples with different spectroscopic qualities $Q=2, 1.5, 1, 0$ as a function of $I_{AB}$ and $K_{AB}$ magnitudes. 

\vskip 0.1 truecm
As table \ref{table:subsamples} shows, there is some correlation between the success rate for a particular subsample, and its median i- and k-band magnitude. The success rate, however, also depends on the redshift distribution in each subsample.
Figure \ref{fig2} shows both the optical ($I_{AB}$) and near-infrared ($K_{AB}$) magnitude distributions of the different spectroscopically identified ($Q=2$) subsamples listed in Table \ref{table:subsamples} as stacked histograms (removing duplications). The high-z subsample (orange) contains the faintest objects, followed by the X-ray subsample (red). Figure \ref{fig3} shows the corresponding redshift distributions. The X-ray detected AGN fill in the "redshift desert" ($1.5<z<3$) known for normal galaxies, thanks to their typically strong emission lines bluer than 3000 \AA. The high-redshift sample completes the range up to $z \leq 6$. 

\begin{figure}
\begin{center}
\includegraphics[angle=0,scale=.21]{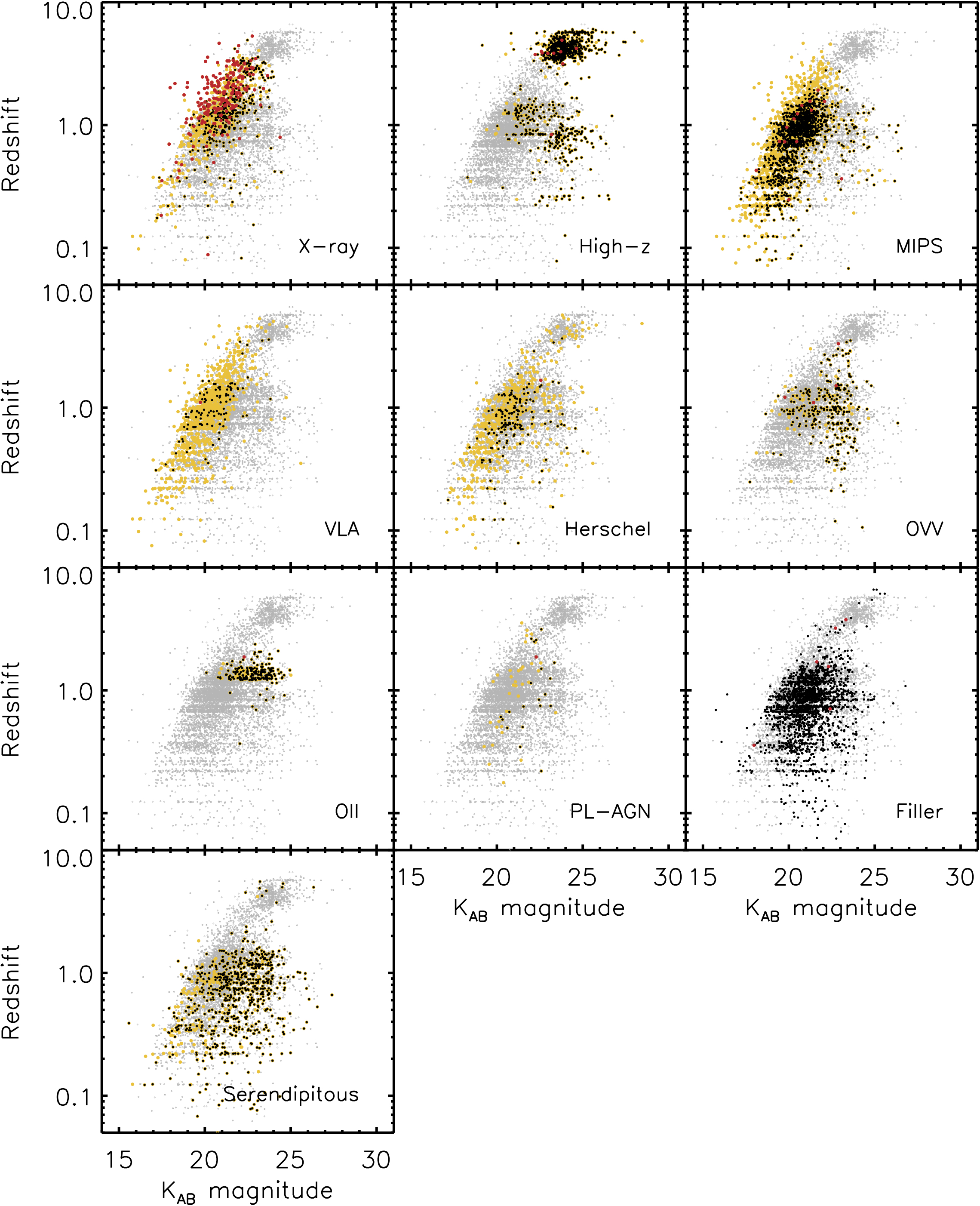}
\caption{Redshift versus $K_{AB}$ magnitude for the different subsamples of objects with $Q\ge1.5$ on the DEIMOS masks. Red asterisks are objects of the particular subsample with broad emission lines. Black points give the unique objects of this subsample, while yellow data points show the objects overlapping with other subsamples. Underlying gray points in each panel show the total DEIMOS catalog. \label{fig4}}
\end{center}
\end{figure}

\vskip 0.1 truecm
Figure \ref{fig4} shows the distribution of redshift versus $K_{AB}$ magnitude. The total DEIMOS sample is shown in grey and each panel shows the sub-type of the selected targets as listed in Table \ref{table:subsamples}. The unique sources are presented in black, while the shared subsample sources are overplotted in yellow. In red we show the sources with broad emission lines, indicating the presence of an AGN. These sources (mainly QSOs) are clearly dominant at higher redshifts in the X-ray subsample, compared to the lower-luminosity AGN at lower redshifts. In X-ray selected AGN samples there is a well known effect of a significant increase in the fraction of broad-line objects with increasing X-ray luminosity. This can be partially understood in terms of the difficulty to detect broad lines of weak AGN against the host galaxy light, but also due to a strong dependence of the obscured AGN fraction on X-ray luminosity \citep[see e.g.][]{2008A&A...490..905H}. The other subsamples contain only very few broad-line objects (the X-ray objects have been removed from all other subsamples). It is interesting to see, how each of the multiwavelength photometric selections corresponds to a characteristic distribution in this graph. The [OII] emission line selection worked well, yielding a narrow redshift range. The high-redshift photometric selection was rather successful, with the majority of the objects at $z>3$, but with a significant fraction of lower-redshift interlopers, possibly associated with mis-identified color breaks and/or photometric noise. The mid- and far-infrared selections (MIPS and Herschel), as well as the VLA radio sources, have a very similar redshift versus magnitude pattern.    

\begin{figure}
\begin{center}
\includegraphics[angle=0,scale=1.25]{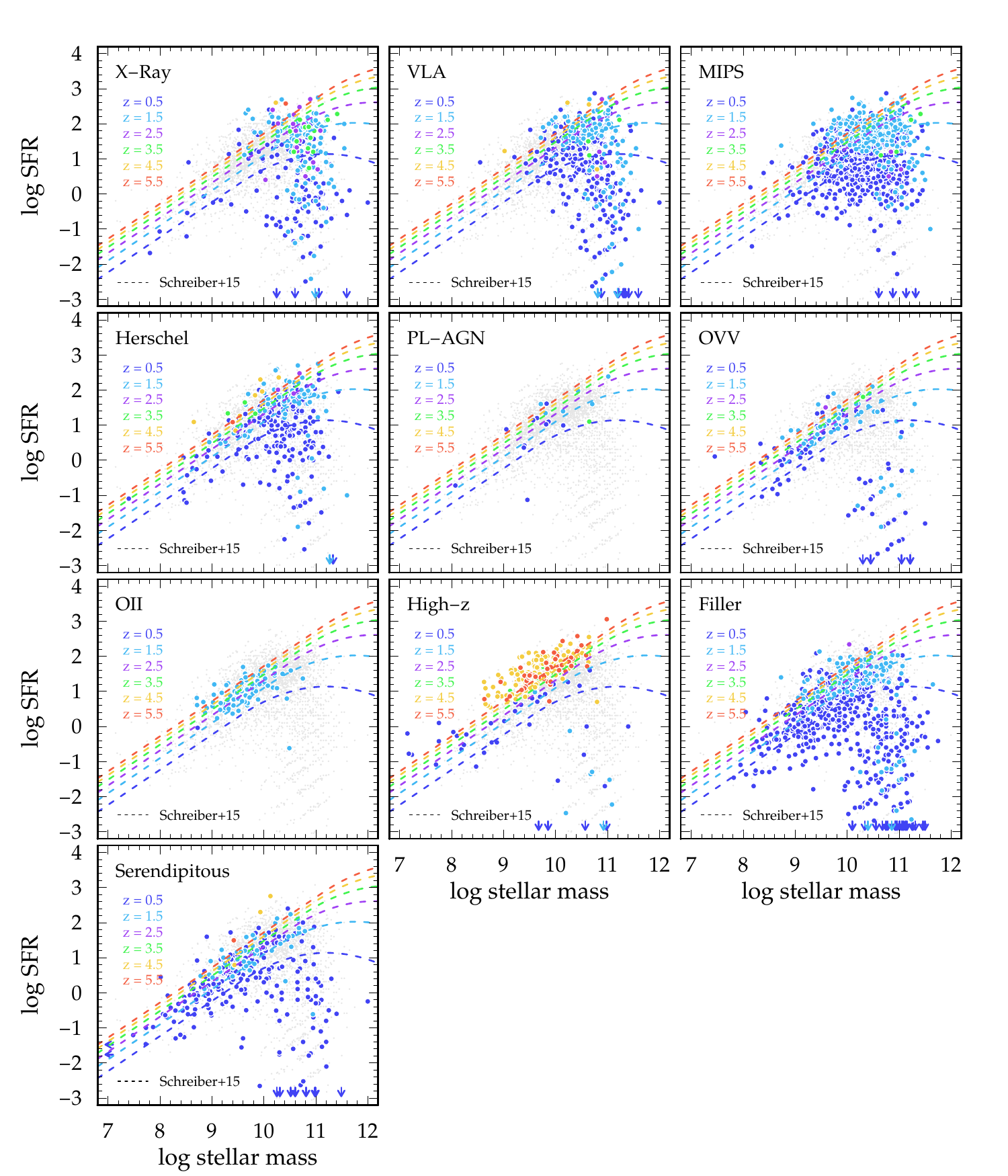}
\caption{Star formation rate versus stellar mass for high-quality spectroscopic and photometric subsamples in the DEIMOS catalog (see text), compared to the star forming main sequence at different redshifts. Dashed lines are derived from Herschel photometry by \cite{2015A&A...575A..74S}. \label{fig5}}
\end{center}
\end{figure}

\vskip 0.1 truecm
Figure \ref{fig5} shows the stellar mass vs. SFR relation for the 10 sub-samples color-coded by their redshifts. The relation for star-forming galaxies at different redshifts (same color code) derived from Herschel photometry by \cite{2015A&A...575A..74S} is shown as dashed line in each panel.
The stellar mass and SFR for our galaxies are obtained by matching the spectroscopic sample to the COSMOS2015 catalog \citep{2016ApJS..224...24L} with a radius of $1.5 \arcsec$. We only consider galaxies with quality flags Q = 2 and further restrict our sample to galaxies for which the photometric and spectroscopic redshifts agree within the uncertainty of the photometric redshifts, i.e., $|z_{spec} - z_{phot}|/(1+z_{spec}) < 0.03$. The latter selection is important to provide the appropriate stellar masses and SFRs for our galaxies since the COSMOS2015 catalog is based on photometric redshifts. In all panels we only use objects without broad emission lines. A full SED fitting using the spectroscopic redshift for all our galaxies is beyond the scope of this paper but will be published in a future work.
We first note that the bulk of galaxies in our sample is consistent with being on the star forming main-sequence. However, the sample also shows a large diversity. In particular, there is a non-negligible fraction of galaxies that are 2-3 magnitudes below the main-sequence at a fixed stellar mass, hence are considered as quiescent in star formation.
Such galaxies are not seen in the power-law AGN (``PL-AGN'') and ``OII'' sub-samples. In the former, the sample size might be too small to detect them. The latter is by definition targeting galaxies with [OII] emission via narrow-band and color-excess selections, thus biased to star-forming systems.
The sub-sample of color and narrow-band selected high-redshift galaxies (``High-z'') contains some contamination (<20\%) from low redshifts (see also Figure 3). Note that the accurate multi-band photometric redshifts in the COSMOS2015 catalog would have picked out these interlopers as opposed to the color and narrow-band selections applied here. About two-thirds of the contaminants are star-forming and about one-third are quiescent galaxies. These could be mistaken as high-redshift galaxies because of their red color caused by dust or old stellar populations or also pure photometric scatter. We checked the spectra of the quiescent galaxies. Almost all of them seem to show strong H+K features, indicating that they are quiescent or in a post-starburst phase.

\begin{figure}
\begin{center}
\includegraphics[angle=0,scale=.64]{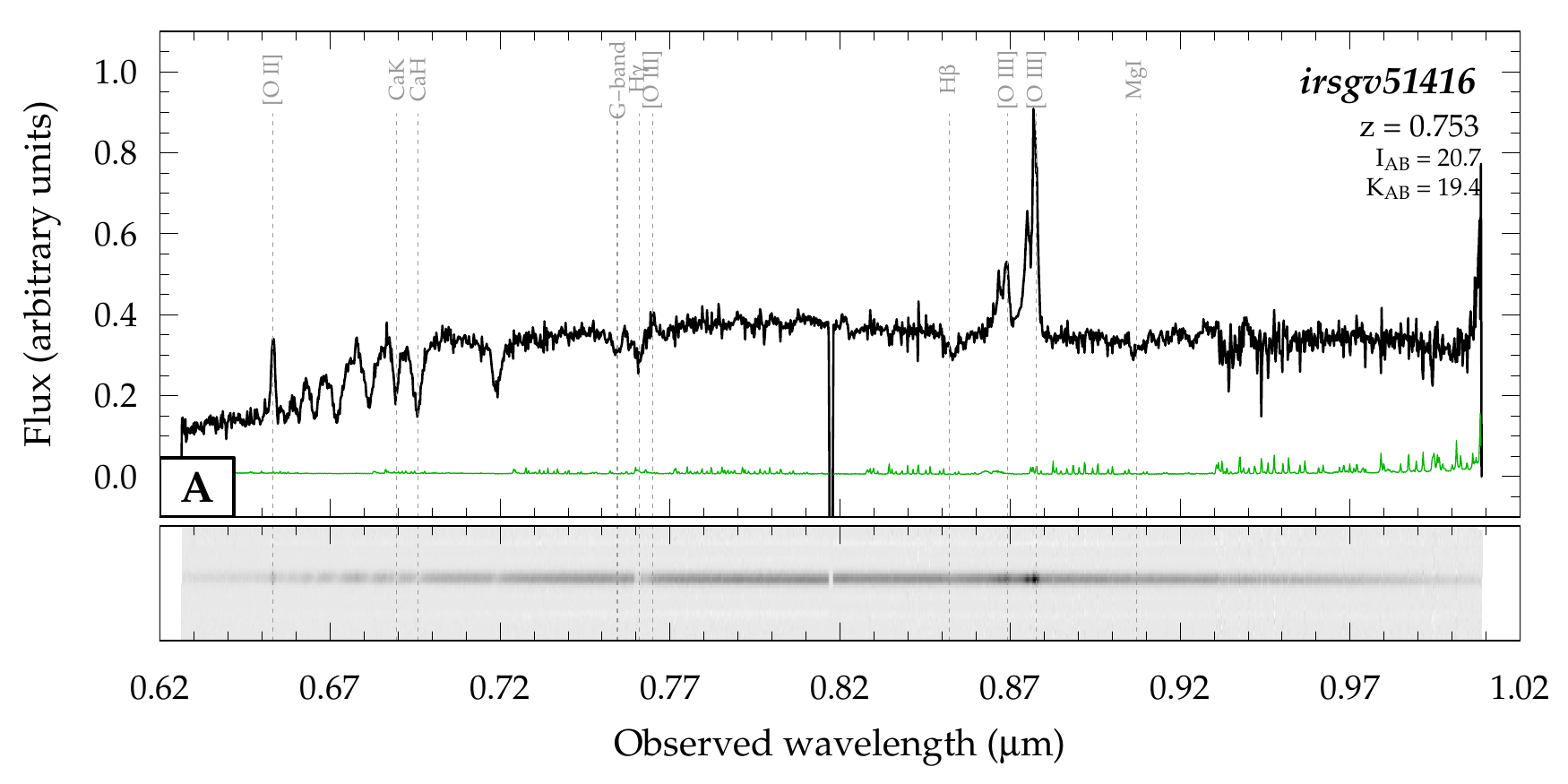}
\includegraphics[angle=0,scale=.64]{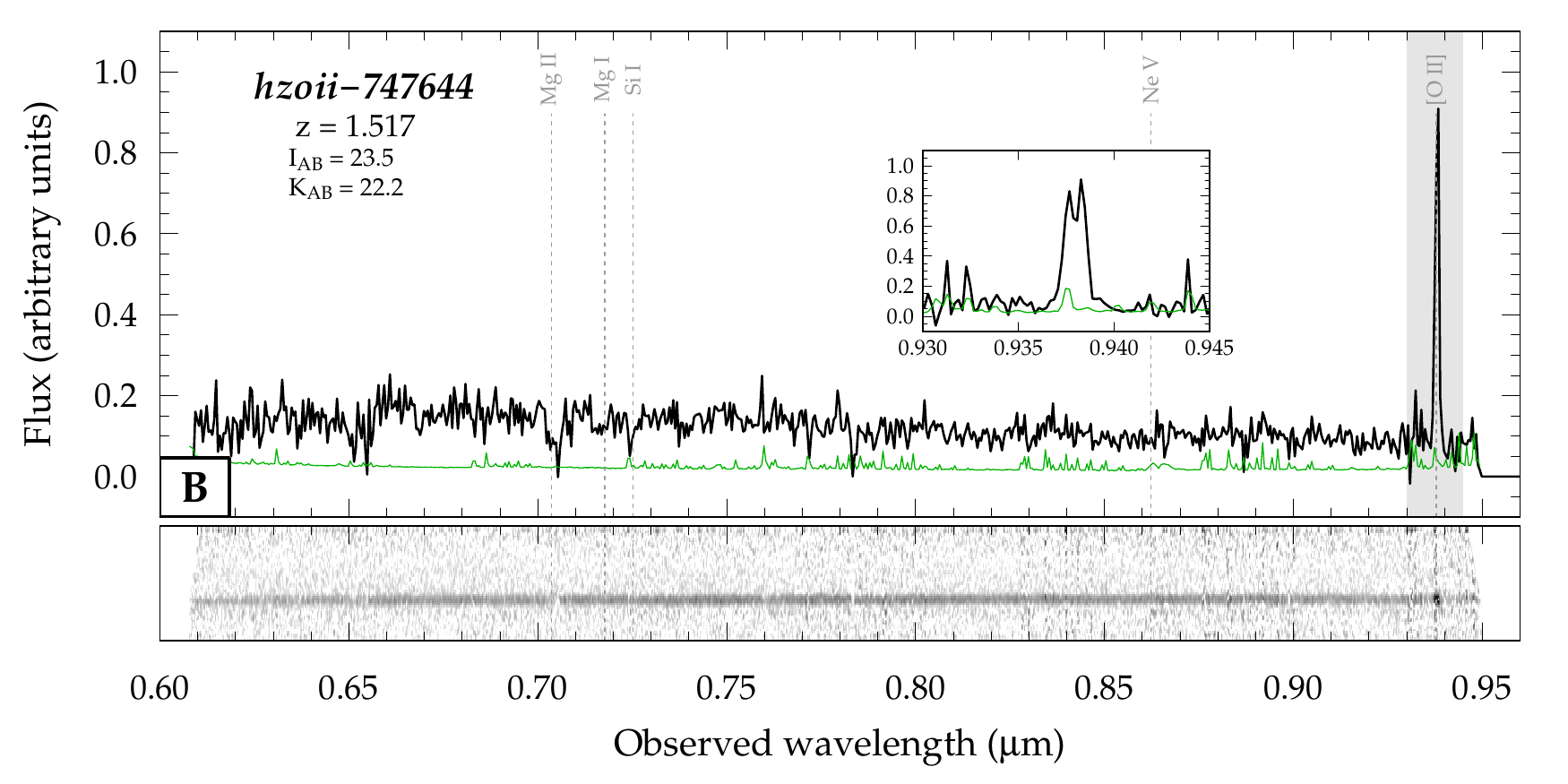}
\includegraphics[angle=0,scale=.64]{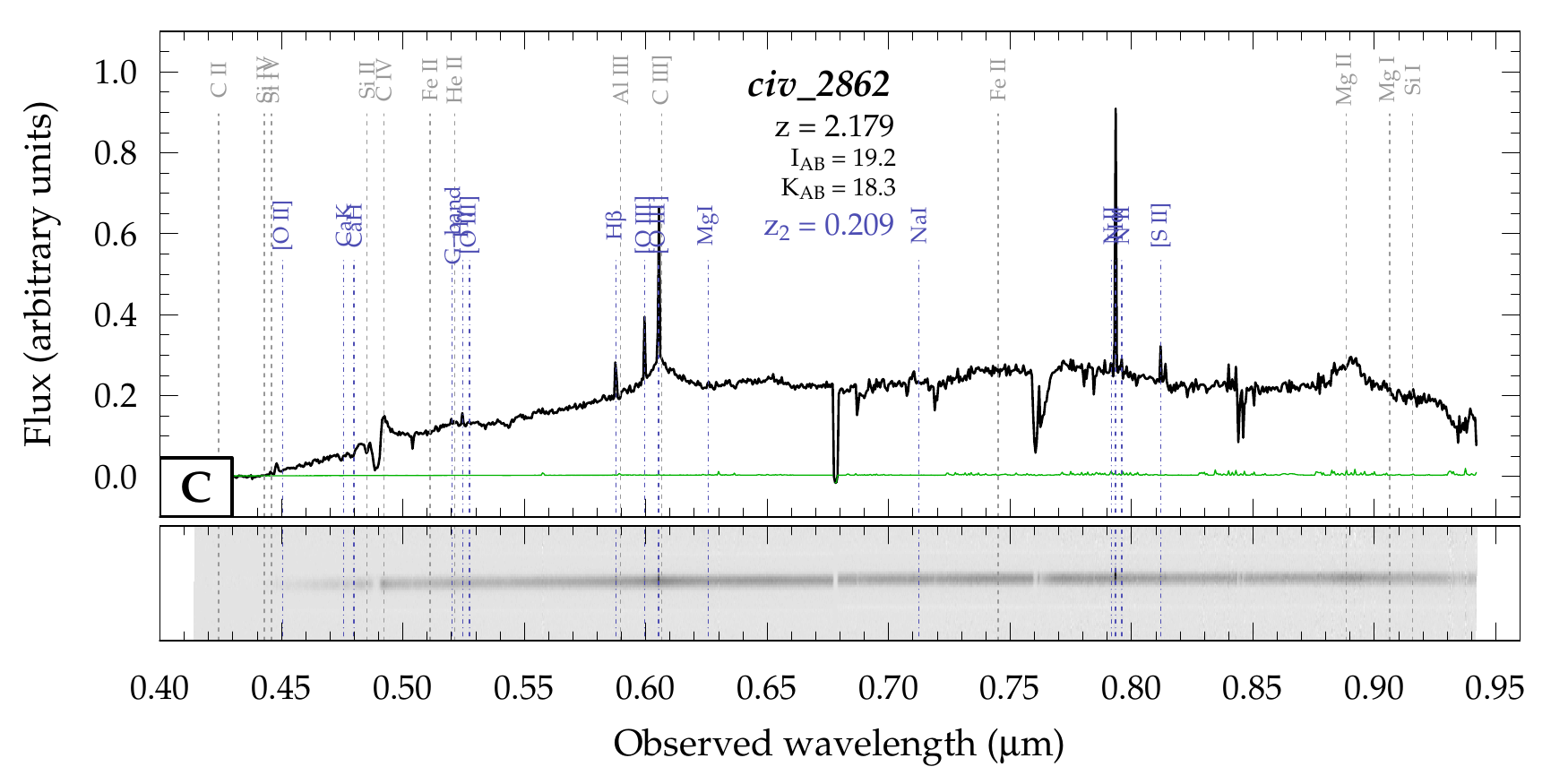}
\includegraphics[angle=0,scale=.64]{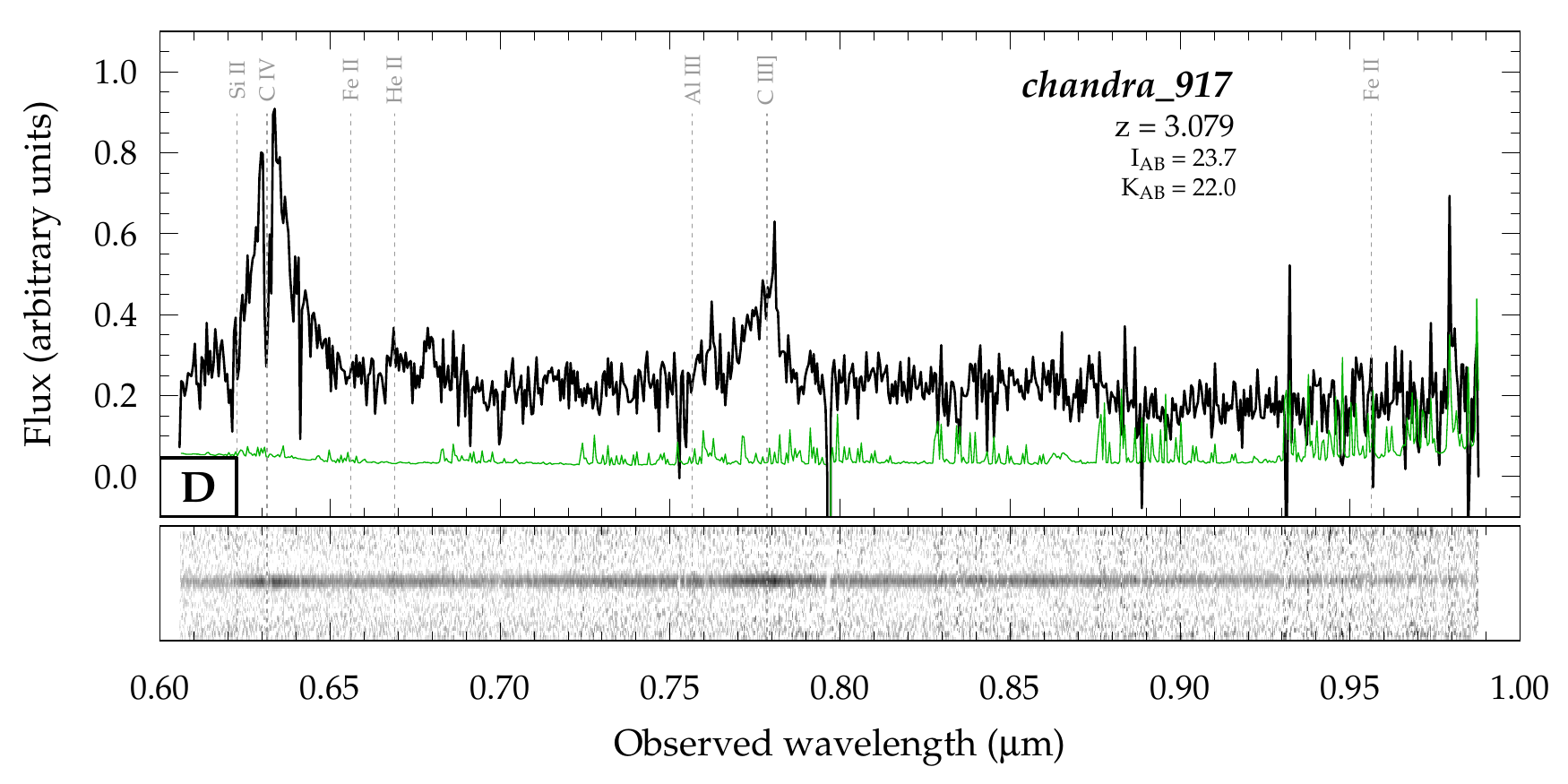}
\caption{Sample of different spectra in the range $0.75<z<3.1$. Fluxes are in arbitrary units. The green line shows the noise level due to the sky background subtraction. \label{fig6}}
\end{center}
\end{figure}

\begin{figure}
\begin{center}
\includegraphics[angle=0,scale=.64]{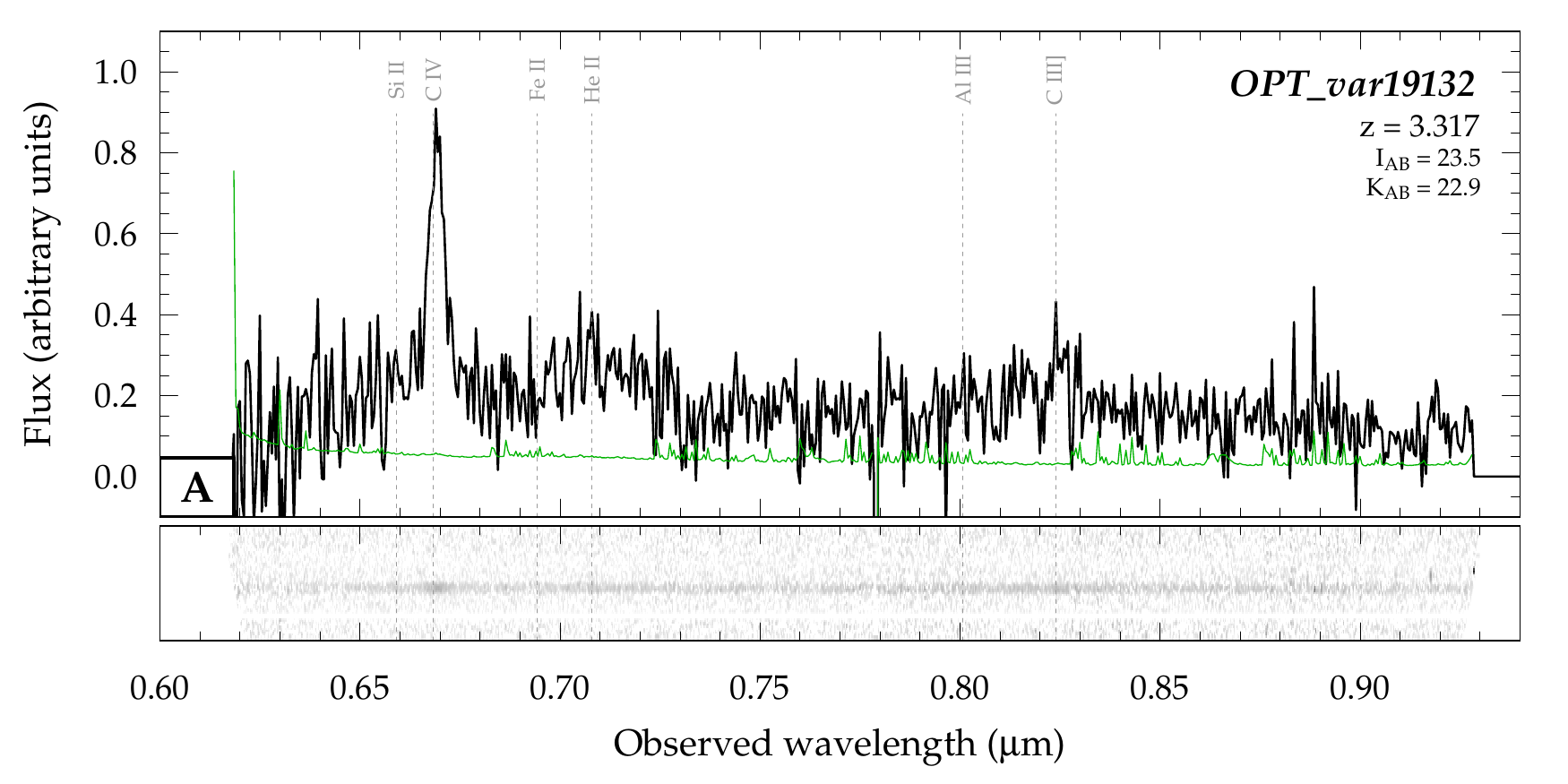}
\includegraphics[angle=0,scale=.64]{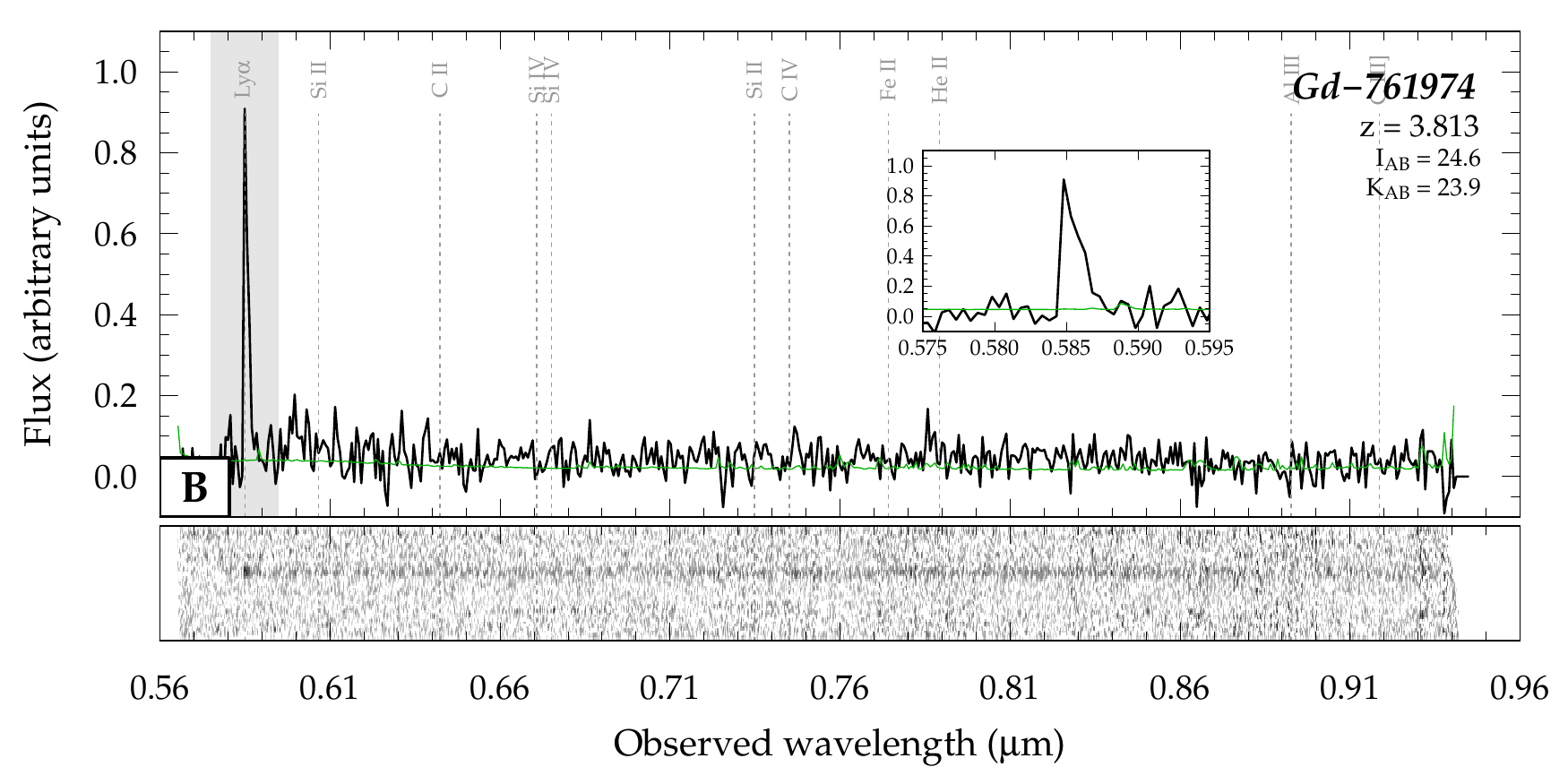}
\includegraphics[angle=0,scale=.64]{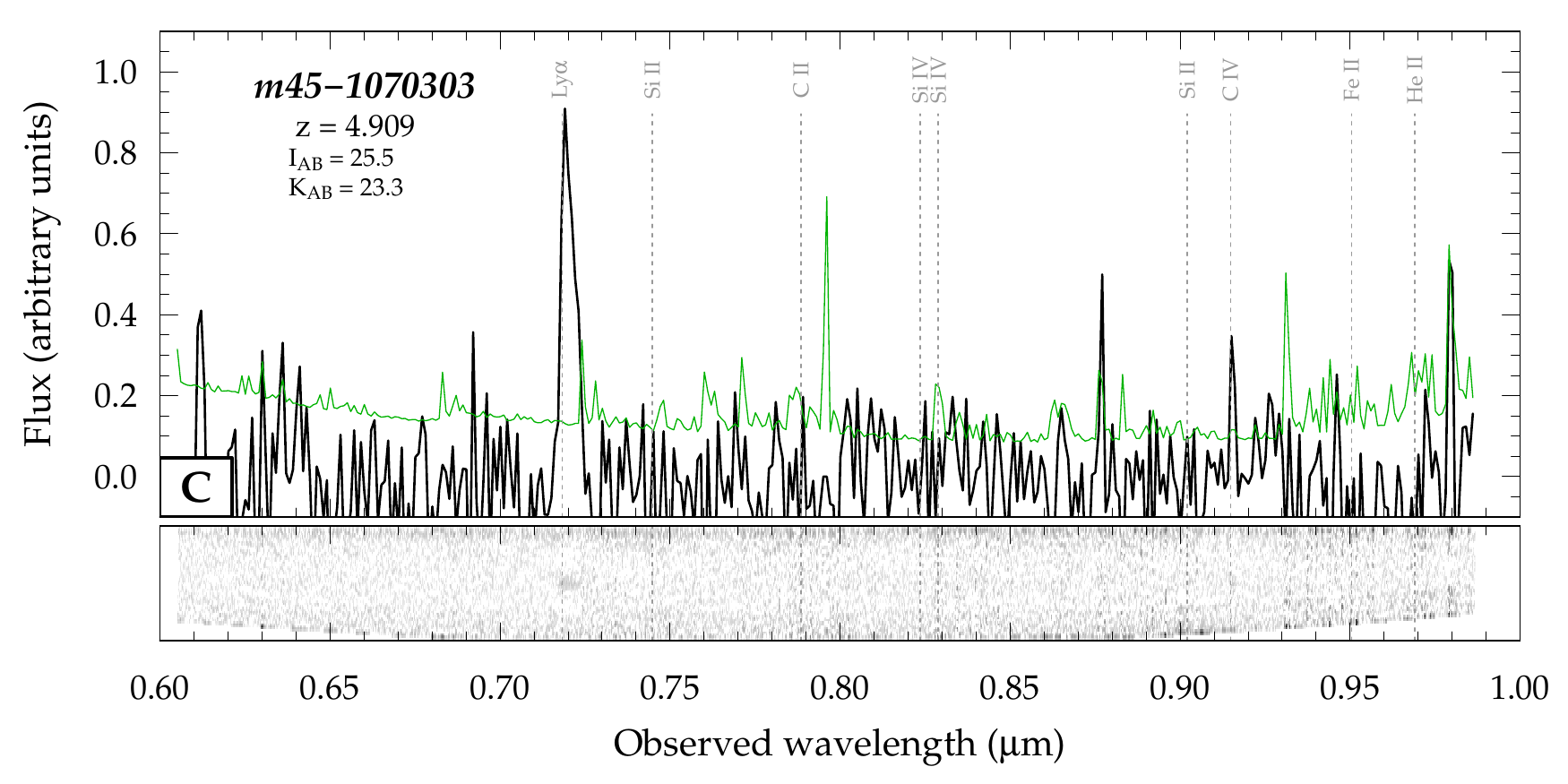}
\includegraphics[angle=0,scale=.64]{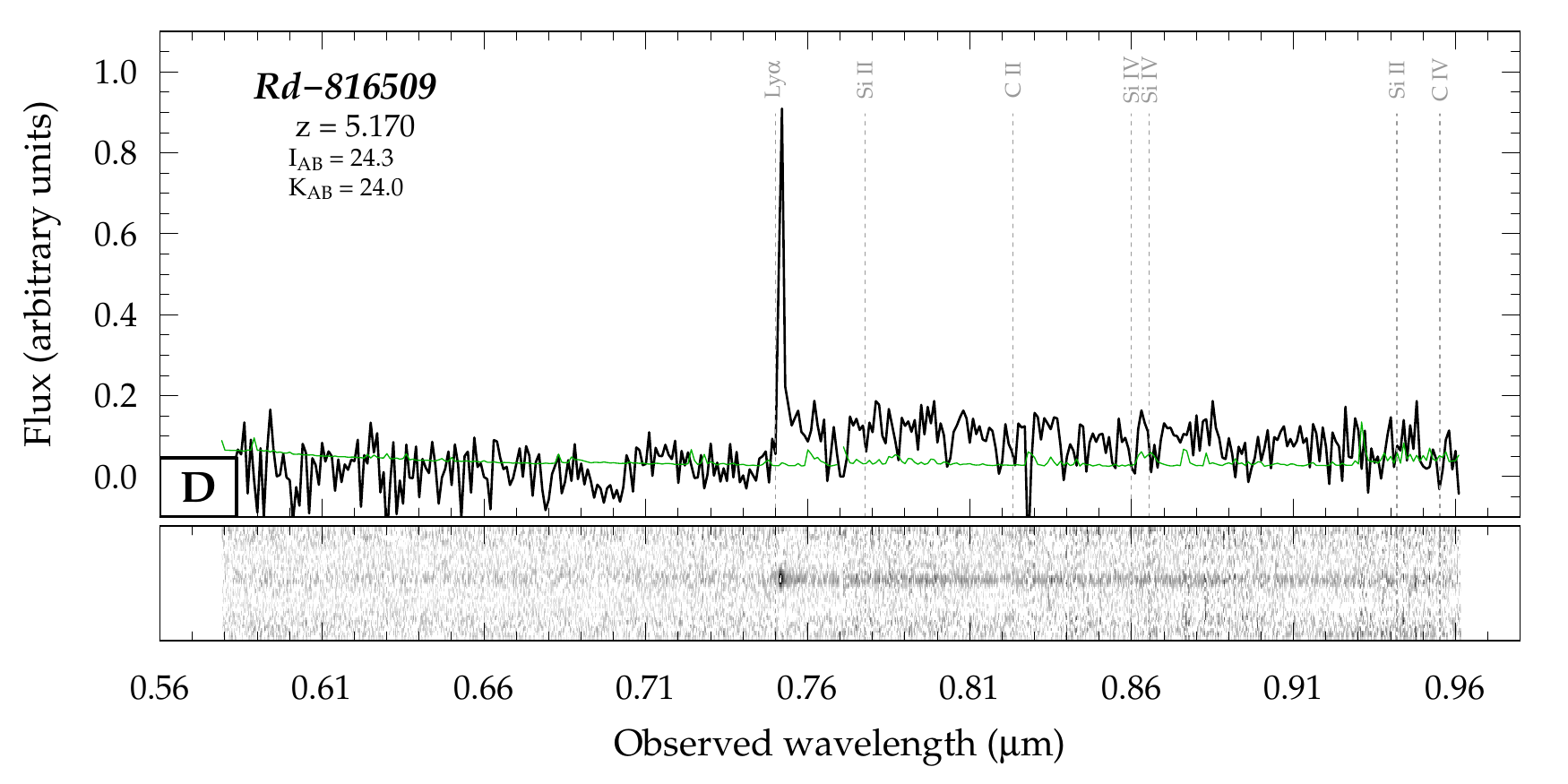}
\caption{Sample of different spectra in the range $3.3<z<5.2$, as in Fig. \ref{fig6} \label{fig7}}
\end{center}
\end{figure}

\vskip 0.1 truecm
Figures \ref{fig6} and \ref{fig7} show some examples of high-quality spectra to illustrate different elements of the spectral classification applied here.

\begin{itemize}

\item The object irsgv51416 (Fig. \ref{fig6}A) is an emission line galaxy at $z=0.753$. It has several interesting features. First, the [OIII] 4960{\AA} and 5008{\AA} doublet emission lines show a broad, double-humped structure, indicating a rapidly rotating gas disk in this galaxy. The difference between the blue and the red wing of the double-humped line profile corresponds to a velocity difference of $770 \ km/s$, or a rotation velocity of $\sim 390 \ km/s$. This is the only object in the DEIMOS sample showing such a rapidly rotating disk. Simultaneously, this object shows strong Ca-H\&K 3940{\AA} and 3960{\AA} absorption lines representative of an old stellar population, together with strong Balmer absorption lines of $H\delta$ 4103{\AA}, $H\epsilon$ 3971{\AA}, $Hf$ 3888{\AA}, etc., corresponding to an A-star population, indicative of more recent, quenched star formation. Together with the relatively weak [OII] 3728{\AA} emission line, this object therefore qualifies as a textbook example of a so called ``E+A'' galaxy. Interestingly, this object is not detected at X-ray, infrared or radio wavelengths. In the whole DEIMOS sample we detected about 30 objects with E+A features, but often together with stronger [OII] 3728{\AA} emission.

\vskip 0.1 truecm
\item The object in Figure \ref{fig6}B, hzoii-747644, has been selected as a strong [OII] 3728{\AA} emission line candidate through its intermediate band filter excess, paired with significant emission in the bluer bands. Indeed, a strong [OII] 3728{\AA} line has been detected in its spectrum, which shows the characteristic double-humped line profile (see inset). In the bluer part of the spectrum we also see significant absorption lines of MgII 2799{\AA}, MgI 2852{\AA}, and SiI 2881{\AA}. The total DEIMOS catalog contains more than 1000 detections of the [OII] 3728{\AA} doublet, which is therefore by far the most abundant emission line in our sample. 

\item The object civ\_2862 in Figure \ref{fig6}C, an X-ray selected broad-line QSO at $z=2.179$, is interesting, because in addition to broad CIV 1549{\AA} , CIII] 1909{\AA} and MgII 2799{\AA} lines, plus significant absorption features, it contains a second spectrum of an emission line object at $z=0.209$. The HST ACS image of this object shows the QSO as a point-like object (actually close to a bright star), and a separate galaxy, which apparently is in the foreground of the QSO. It is possible, that the QSO is slightly gravitationally lensed by the galaxy, but there are no other lensed images of the QSO. This object is one of the best examples of a superposition of spectra of two objects in the same slit. We have found quite a number of double spectra in our sample, which will be discussed in section \ref{secdual}.

\item Object chandra\_917 in Figure \ref{fig6}D is an example of an standard X-ray selected broad-line QSO at $z=3.097$, with some absorption features superposed on the broad lines. Figure \ref{fig7}A shows another broad-line object at $z=3.317$, in this case selected through its optical variability. 

\item Figure \ref{fig7}B-D shows \Ly \ emitters selected from our high-z samples. Although often only a single strong \Ly \ 1216{\AA} \ line is visible, the line can be identified through the characteristic skewed emission line shape \citep[see e.g.][]{2010PhDT.......254L} caused by the neutral intergalactic medium at high redshifts (see insert in Figure \ref{fig7}B). In the case of the object Rd-816509 we also clearly see the continuum drop blueward of the \Ly \ line. 

\end{itemize}

\section{The DEIMOS Redshift catalog}

Table \ref{table:objects} shows the DEIMOS spectral identifications and redshifts in the COSMOS field. The full catalog can be accessed online on the official COSMOS webpage\footnote{\url{http://cosmos.astro.caltech.edu}} and through the Vizier service. The first column gives an object identifier from the major two photometric catalogues. An "L" in front of the number refers to the red multiband-band selected catalogue of \cite{2016ApJS..224...24L}. A "C" in front of the number refers to the i-band selected catalogue of \cite{2007ApJS..172...99C} and \cite{2009ApJ...690.1236I}.\footnote{\url{http://cosmos.astro.caltech.edu/page/photom}} If an object is not present in either of these catalogs it does not have an identifier. Columns two and three give the J2000 Right Ascension and Declination for each object. The fourth column gives an identifier, in which subsample a particular object is present. It is a decimal representation of a binary flag\footnote{sel=512*X+256*hiz+128*M+64*VLA+32*H+16*OVV+8*OII+4*PL+2*Fil+1*ser} containing the X-ray, high-z, MIPS, VLA, Herschel, OVV, OII, PL AGN, Filler and Serendipitous flag following the order in table \ref{table:samples} and figure \ref{fig0}. The fith and sixth columns give the $I_{AB}$ and $K_{AB}$ AB magnitudes, based on the ultradeep Subaru Hyper Suprime-Cam \citep{2017arXiv170600566T} and UltraVista \citep{2016ApJS..224...24L}, the Subaru Suprime-Cam \citep{2009ApJ...690.1236I}, and the Hubble ACS \citep{2007ApJS..172..196K} photometric catalogs. Because of field-coverage, bright star cut-outs, blending or other confusion issues not all objects in the spectroscopic catalogue are covered by a single photometric catalogue, and we thus have to refer to various different imaging datasets. The seventh column gives the spectroscopic redshift $z$. The eigth column gives the spectroscopic quality flag ($Q_f=0, 1, 2, 3, 4, 9, 11, 12, 13, 14, 19$), following the original zCOSMOS scheme \citep{2009ApJS..184..218L}, where values 11-19 indicate broad emission lines (see above). The ninth column gives the reduced spectral quality flag (see above). Finally, the tenth column gives remarks for most objects, in particular indicating the spectral features detected, e.g. the \Ly \ and Balmer lines (\Ha, \Hb, \Hg, ...) of hydrogen, or the MgII line, as well as the [CIV], CIII], CII], [OII], [OIII], NII, and [SII] emission lines. A ``d'' behind an emission line designation indicates a detected line doublet. A ``br'' behind an emission line refers to a broad emission line profile. An ``abs'' behind a line indicates its appearance in absorption rather than emission. H\&K and G correspond to the Ca-H 3940{\AA} \ and Ca-K 3960{\AA} \ absorption lines and the G 4304{\AA} \ absorption band, respectively. Other prominent absorption lines are MgI 5175 {\AA} \ and NaI 5892 {\AA} \. Finally, ``E+A'' features indicate the forest of spectral emission and absorption features (``ringing'') between the [OII] line and Ca-H \& K, characteristic of post-starburst (E+A) galaxies. 

% TABL% TABLE 3:  Spectral Identifications
\scriptsize
%long table for AAStex
\startlongtable
%switch to this for emulateapj
%\LongTables
\begin{deluxetable*}{lllccccccl}
% \tablewidth{0pt}
\tablecaption{List of spectral identifications}
\tablehead{
\colhead{ID} &
\colhead{R.A.} &
\colhead{Dec.} &
\colhead{Sel} &
\colhead{i$_{AB}$} &
\colhead{k$_{AB}$} &
\colhead{Redshift} &
\colhead{$Q_f$} &
\colhead{Q} &
\colhead{Remarks} \\
\colhead{} &
\colhead{(J2000)} &
\colhead{(J2000)} &
\colhead{} &
\colhead{(Mag)} &
\colhead{(Mag)} &
\colhead{} &
\colhead{} &
\colhead{} &
\colhead{} }
\startdata
C1785365 & 149.358553 & 2.750301 & 512 & 20.11 & 20.21 & 0.5655 & 2 & 1.5 & H+K? \\
C1343700 & 149.366394 & 2.37669396 & 128 & 22.89 & 21.61 & 0.708 & 3 & 2 & [OII],[OIII] \\
C1347270 & 149.3690949 & 2.34408808 & 128 & 21.73 & 20.14 & 0.708 & 3 & 2 & [OIII]d \\
C1784226 & 149.37375 & 2.7765083 & 2 & 18.17 & 17.16 & 0 & 4 & 2 & M-star \\
C1785765 & 149.3740997 & 2.7369051 & 2 & 21.37 & 19.63 & 0.549 & 2 & 1.5 & [OII],H+K? \\
C1344930 & 149.3744965 & 2.41525006 & 2 & 18.36 & 15.95 & 0 & 4 & 2 & star \\
C1344754 & 149.3755035 & 2.42347503 & 128 & 24.28 & 21.75 & 1.079 & 1 & 1.5 & \Hg,[OIII]? \\
C1346553 & 149.377594 & 2.36321092 & 2 & 18.49 & 18.19 & 0 & 4 & 2 & star \\
C1346616 & 149.3811035 & 2.36196303 & 512 & 21.84 & 20.52 & 0.772 & 4 & 2 & [OII],[OIII]d \\
C1345095 & 149.3820953 & 2.40950799 & 128 & 22.5 & 21.04 & 0.93 & 3 & 2 & [OII],\Hb \\
C1345098 & 149.38211 & 2.4109753 & 1 & 17.39 & 16.55 & 0 & 22 & 1.5 & star \\
L709816 & 149.3825989 & 2.37418294 & 512 & 22.82 & 21.73 & 3.367 & 14 & 2 & \Ly,CIV(br),CIII] \\
L941303 & 149.386634 & 2.743066 & 512 & 21.62 & 19.26 & 0.737 & 4 & 2 & [OII],H+K,\Hb,MgI \\
L738187 & 149.3869019 & 2.41814995 & 128 & 23.37 & 21.97 & 1.099 & 3 & 2 & [OII]d,K \\
...\\
L365372 & 150.79555 & 1.8459245 & 1 & 22.53 & 24.45 & 0.842 & 24 & 2 & [OII],H+K,\Hb,[OIII]d \\
C1796700 & 150.7964783 & 2.9122829 & 128 & 21.1 & 23.56 & 0.298 & 4 & 2 & \Hb,[OIII]d,\Ha,NII,[SII] \\
C1797283 & 150.8010864 & 2.9061069 & 128 & 20.86 & 23.55 & 0.925 & 4 & 2 & [OII],H+K,\Hb,[OIII]? \\
C1793829 & 150.8035583 & 2.9437439 & 256 & 25.67 &  &  & 0 & 0 &  \\
C1795460 & 150.8054047 & 2.925 & 256 & 25.59 &  & 1.19 & 2 & 1.5 & [OII],H,G?  \\
C1797605 & 150.8085327 & 2.9029269 & 128 & 19.85 & 23.55 & 0.365 & 14 & 2 & \Hb(br),[OIII]d,\Ha(br)NII,[SII]\\
L1041939 & 150.8104248 & 2.8838029 & 128 & 23.16 & 23.55 & 1.203 & 2 & 1.5 & [OII]        \\
C1798722 & 150.8167114 & 2.8923731 & 128 & 20.97 & 23.55 & 0.924 & 4 & 2 & [OII],\Hg,\Hb?        \\
C1798723 & 150.817749 & 2.89088893 & 129 & 19.82 & 23.55 & 0.249 & 24 & 2 & \Hb,[OIII]d,\Ha,NII,[SII] \\
L1046705 & 150.8190918 & 2.8908889 & 8 & 23.7 & 23.55 & 1.352 & 1 & 1.5 & [OII]?        \\
C1797314 & 150.8242493 & 2.9058919 & 128 & 21.09 & 23.55 & 0.918 & 4 & 2 & [OII],K,\Hg,\Hb,[OIII]        \\
L1038944 & 150.8267365 & 2.879705 & 128 & 23.48 & 23.55 &  & 0 & 0 &  \\
         & 150.8373566 & 2.930975 & 256 &  &  &  & 0 & 0 &  \\
C1794871 & 150.8376007 & 2.9305582 & 1 & 23.92 & 23.55 & 1.365 & 3 & 2 & [OII]d \\
\enddata
\label{table:objects}
\tablecomments{The full catalogue can be retrieved from \url{http://cosmos.astro.caltech.edu}}
\end{deluxetable*}
\normalsize

\section{Comparison between redshift surveys}
There is a large variety of spectroscopic surveys in the COSMOS field shown in table \ref{table:samples}, which used different instruments and spectral resolution, different wavebands and target selections, as well as target redshift ranges. The multitude of COSMOS surveys therefore provides a unique opportunity to compare independent redshift measurements, and to cross-calibrate their respective individual quality assessments. For this purpose we have used our internal master catalog where we have collected all data available in the literature (Salvato M., in prep.). In table \ref{table:samples}, the quality entries in the individual surveys have been reduced to four basic quality classes, i.e. $Q_f=1-4$, as well as the 9 for single line spectra. For those surveys with a more elaborate quality flag (e.g zBRIGHT, zDEEP, VUDS, DEIMOS) the integer value of the $Q_f$ quality flag has been taken modulo 10 for the purpose of this comparison (e.g. $Q_f=14.5, 24.5$ etc. were reduced to 4). Objects in each individual survey and quality class was then compared to the overlapping objects in the DEIMOS $Q_f=4$ class. In general the overlap between different surveys should be small, because there is some degree of coordination between the different redshift surveys. However, because of the long time interval covered by our observations, because not all surveys were coordinated, and because there are serendipitous or filler targets, which can overlap with other surveys, there is a sizable sample of objects with spectra in more than one survey. Here we restrict the comparison only to surveys, where there is a meaningful statistical sample ($>15$) of overlapping objects. 
Because two imperfect samples are compared to each other in each case, this procedure only provides information on the combined accuracy of the two samples in question. The true accuracy of the individual sample should be somewhat better. 
Table \ref{table:samples} shows the results of this analysis. The best spectroscopic surveys in comparison to the highest quality objects in our DEIMOS sample, with less than 2.5\% spectroscopic failures and redshift difference less than $1.5\times10^{-3}$ after removing the outliers, are the highest quality class ($Q_f=4$) for DEIMOS-C3R2, FMOS-14, FOCAS, FORS2-11, Gemini-S, LRIS, SDSS, zBRIGHT, and zDEEP. The quality comparison for the DEIMOS sample itself was done only against these highest quality surveys ($Q_f=4$). There is an overlapping sample of 568 DEIMOS objects with the combination of $Q_f=4$ objects from DEIMOS-C3R2, FMOS-14, FOCAS, FORS2-11, Gemini-S, LRIS, SDSS, zBRIGHT, and zDEEP. Differences in quality are partially due to the distribution of redshifts and magnitudes and telescope/instruments involved. However, the time consuming visual inspection and vetting of every redshift by several independent collaborators also plays an important role in the highest-quality surveys. The DEIMOS survey accuracy is among the best spectroscopic samples in the COSMOS field, with less than 2\% spectroscopic failures and a redshift accuracy better than $10^{-3}$.  

% TABLE 2:  Comparison 
% \scriptsize
\begin{deluxetable*}{lcccccc}
% \tablewidth{0pt}
\tablecaption{Comparison with other spectroscopic surveys}
\tablehead{
\colhead{Sample} &
\colhead{Total} &
\colhead{$Q_f$=4} &
\colhead{$Q_f$=3} &
\colhead{$Q_f$=2} &
\colhead{$Q_f$=1} &
\colhead{$Q_f$=9}}

\startdata
3D-HST       & 69/0.0/3.4    & 69/0.0/3.4 \\
DEIMOS-C3R2  & 38/2.6/1.1    & 31/0.0/1.1 & 5/0.0/0.2    & 1/0.0/0.0     & 1/100/- \\
FMOS-14      & 12/0.0/0.2    & 11/0.0/0.2 &               & 1/0.0/0.4     \\
FMOS-15      & 155/12.9/7.4   & 62/1.6/2.8  & 50/6.0/6.7  & 11/63.6/1.3   & 32/28.1/13.3 \\
FMOS-16      & 70/4.3/7.5    & 70/4.3/7.5 \\
FOCAS        & 12/0.0/1.0    & 12/0.0/1.0 \\
FORS2-11     & 41/2.4/0.6    & 41/2.4/0.6 \\
FORS2-15     & 140/20.0/11.0 & 54/3.7/5.5  &               & 21/52.4/14.0  & 9/100/-      & 52/11.5/11.7 \\
Gemini-S     & 29/3.4/0.9    & 19/0.0/0.2 & 9/11.1/1.6   & 1/100/-       \\
IMACS        & 194/33.0/9.8  & 114/21.9/8.3 & 17/41.2/1.4 & 39/48.7/12.2  & 24/54.2/15.5 \\
IRS          & 22/0.0/7.5     & 15/0.0/3.8  & 5/0/0.5     &               & 2/0.0/22.4   \\
LRIS         & 45/11.1/1.5   & 13/0.0/1.5  & 8/0.0/2.3   & 8/12.5/1.5   & 16/25.0/1.1 \\
MMT          & 35/0.0/4.3     & 35/0.0/4.3  \\
MOSFIRE MOSDEF & 12/16.7/1.6  & 12/16.7/1.6 \\  
PRIMUS       & 1657/18.4/10.9 & 824/4.6/5.8 & 309/27.8/11.5 & 437/41.2/14.0 \\
SDSS-DR14    & 82/2.4/1.0     & 82/2.4/1.0  \\
VIMOS LEGA-C & 198/1.0/3.2    & 198/1.0/3.2 \\
WFC3 grism   & 20/0.0/9.0     & 10/0.0/2.4& 9/0.0/4.7 & 1/0.0/39.6\\  
zBRIGHT      & 1146/5.9/4.6   & 268/2.2/0.7 & 413/1.7/2.6 & 229/5.7/3.8 & 134/29.1/12.1 & 79/3.8/4.6 \\
zDEEP        & 112/40.2/2.4   & 28/0.0/1.2  & 21/9.5/1.0  & 19/52.6/2.3   & 40/77.5/5.4 & 4/50.0/1.6\\
{\bf DEIMOS} &{\bf 568/3.4/2.7}&{\bf 493/1.8/0.9}& {\bf 29/17.2/6.4}&{\bf 19/15.8/2.7}&{\bf 13/7.7/12.8}&{\bf 8/12.5/3.0} \\
\enddata
\label{table:samples}
\tablecomments{$Q_f$ is the quality class in each of the spectral surveys (see text). The three entries in each cell (1/2/3) are (1) total number of objects, (2) percentage of outliers ($\delta z/(1+z)>0.05$), and (3) redshift accuracy $<\delta z/(1+z)>$ in units of $10^{-3}$. References for the spectroscopic surveys are: 
3D-HST \citep{2016ApJS..225...27M},
DEIMOS-C3R2 \citep{2017ApJ...841..111M},
FMOS-15 \citep{2015ApJ...806L..35K},
FMOS-16 (T. Nagao, priv. comm),
FORS2-11 \citep{2011ApJ...742..125G},
FORS2-15 \citep{2015AA...575A..40CC},
Gemini-S \citep{2014MNRAS.443.2679B},
IMACS \citep{2009ApJ...696.1195T},
LRIS (C. Casey, priv. comm.),
MMT \citep{2006ApJ...644..100P},
PRIMUS \citep{2011ApJ...741....8C}, 
SDSS-DR14 {\url{http://www.sdss.org/dr14/data\_access/}}, 
VIMOS-LEGA-C \citep{2016ApJS..223...29V},
zBRIGHT \citep{2009ApJS..184..218L}, 
zDEEP (Lilly et al., priv. comm).
}

\end{deluxetable*}
% \normalsize

\section{Protoclusters and large-scale structures}

Galaxies in the centers of nearby rich clusters and groups are passive, with little or no ongoing star formation \citep[e.g.][]{2006MNRAS.373..469B}. Models for the evolution of cluster galaxies at low redshift based on the studies of cluster galaxy colors and luminosity functions point to galaxy populations that had a violent phase of star formation at high redshifts ($z>2$) and are passively evolving ever since \citep{2010ApJ...720..284M}. On the other hand, actively star forming galaxies, luminous and ultraluminous infrared galaxies, as well as AGN are typically found in the outskirts of nearby clusters. A statistical study of clusters in the redshift range $1<z<1.5$ shows that the fraction of star forming galaxies is systematically suppressed in the cluster centers and increases to the value of the field galaxies towards the cluster outskirts \citep{2013ApJ...779..138B}. This is true for clusters at redshifts $z<1.4$, while at higher redshifts the trend reverses and the fraction of star forming galaxies increases significantly towards the cluster center to values above the field galaxy ratio \citep{2010ApJ...721..193P,2013ApJS..206....3S,2016ApJ...825..113D}. 

\begin{figure}
\begin{center}
\includegraphics[angle=0,scale=.55]{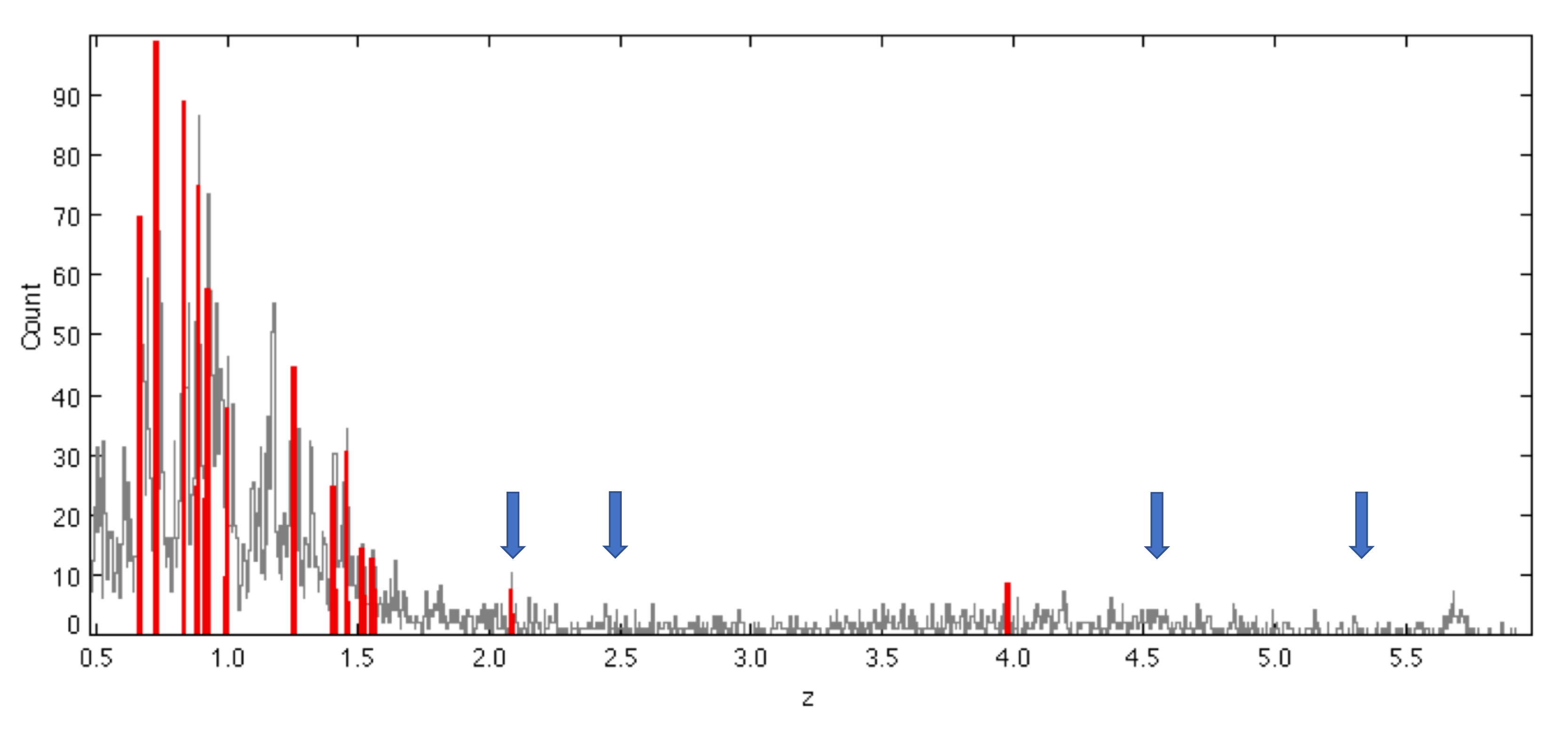}
\caption{Redshift distribution of the catalog objects in bins of $\Delta z=0.005$. Thirteen redshift spikes in the range $0.6<z<4.0$ have been colored in red. Blue arrows indicate previously reported proto-clusters discussed in the text. \label{fig8}}
\end{center}
\end{figure}

The highest redshift bona fide cluster of galaxies has been discovered at $z\sim2.5$ \citep{2016ApJ...828...56W} with a violently starbursting core. At $z>1.4$ there are about 5 known clusters, of which two have masses above $10^{14} M_\odot$ \citep{2010ApJ...716.1503P, 2016MNRAS.463.3582M}. Massive clusters at higher redshifts are extremely rare, because only very few regions in the universe had time to virialize by that cosmic time. Cosmological simulations including cold dark matter and baryons \citep[e.g.][]{2005Natur.435..629S} suggest that most massive clusters of galaxies have started their life at high redshift as over-densities of massive galaxies - ``proto-clusters'' - which have not had time to virialize and therefore occupy regions in space about two orders of magnitudes larger than local clusters. We therefore have to look at proto-clusters \cite[e.g.][]{2015Natur.522..455C} to study the early violent formation phase of the progenitors of today's cluster galaxies. 

\vskip 0.1 truecm
Significant spikes in the redshift distribution of X-ray selected samples have first been identified at $z\leq1.1$ in the Chandra Deep Field North \citep{2002AJ....124.1839B} and South \citep{2002astro.ph..2430H,2003ApJ...592..721G,2004ApJS..155..271S}, but are also present in the photometric redshift distribution of the X-ray sources in the Lockman Hole \citep[e.g.,][]{2012ApJS..198....1F}, XMM-XXL \citep{2017MNRAS.469.3232G} and STRIPE82X \citep{2017ApJ...850...66A}. They point to AGN in superclusters or sheet-like structures of the cosmic web. The question, whether AGN activity or star formation is enhanced in these structures compared to the field, remains open. With the advent of the large photometric and spectroscopic surveys in the COSMOS field, a number of rich proto-cluster structures have been identified at higher redshifts, which are believed to be the progenitors of some of the most massive clusters of galaxies in the local Universe. One of the more prominent of those is a massive proto-cluster of galaxies at a redshift of $z\sim5.3$, with a size of more than $13\,{\rm Mpc}$ and containing a luminous quasar as well as a galaxy with a large amount of molecular gas \citep{2011Natur.470..233C}. Another massive proto-cluster in the COSMOS field was recently discovered at $z=4.57$ with $\log(M_{h}/M_{\odot})_{z=0}\sim14.5-15$ by \cite{2017arXiv170310170L}. A third interesting object is a massive, distant proto-cluster at $z=2.47$, found serendipitously during a spectroscopic redshift survey of dusty star-forming galaxies detected by Scuba-2 in the COSMOS field. This structure may be seen in a phase of violent star formation \citep{2015ApJ...808L..33C}, and may be connected to a large overdensity of \Ly\ emitters found in the HETDEX pilot survey \citep{2011ApJS..192....5A} of the COSMOS field. Finally, a large-scale structure around  $z\sim2.2$ discovered in the Z-FOURGE photometric redshift survey in COSMOS \citep{2012ApJ...748L..21S}, could be confirmed spectroscopically as a Virgo-like cluster ancestor at z = 2.095 \citep{2014ApJ...795L..20Y}. Both the $z=2.47$ and the $z=2.09$ feature are also present in the list of 36 candidate 15 Mpc-scale protocluster structures identified at redshifts $z=1.6-3.1$ in the COSMOS field through photometric redshifts \citep{2014ApJ...782L...3C}. \cite{2013ApJ...765..109D} used the zDEEP sample to search for groups of galaxies in the COSMOS field in the redshift range $1.8<z<3$ within a physical distance of 500 Mpc and a velocity difference of 700 km/s. They identified 42 candidate groups with 3-5 members, and a comparison to mock catalogs indicates that most of them should be in large-scale structures, which later may merge into single groups, but almost none of them should already be virialized. 

\begin{figure}
\begin{center}
\includegraphics[angle=0,scale=.71]{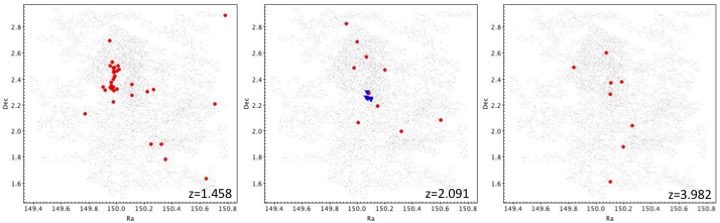}
\caption{Spatial distribution of the objects in the redshift spikes at $z=1.458$ (left), $z=2.091$ (middle), and $z=3.982$ (right). The centers of large galaxy overdensities at $z=2.095$ identified by \cite{2014ApJ...795L..20Y} are shown by blue triangles in the middle panel. Grey points show all objects in the DEIMOS redshift catalog.\label{fig9}}
\end{center}
\end{figure}

\vskip 0.1 truecm
We analyzed our sample of spectroscopic redshifts to look for potential protocluster targets for future follow up. Since the spectroscopic sample is very heterogeneous, it is not possible to perform a rigorous treatment.
Figure \ref{fig8} shows the redshift distribution of the DEIMOS sample in bins of $\Delta$z=0.005. These redshift bins on one hand are about 5 times larger than the intrinsic redshift uncertainty in the catalog, and on the other hand correspond to velocity differences between 630 km/s (at $z=0.8$) and 300 km/s (at z=4), appropriate for the selection of groups in large scale structures (see above). Several significant redshift spikes are seen in this figure at redshift $z<1.3$, corresponding to large-scale structure identified previously in deep fields (see above), and also in the COSMOS field \citep[see e.g.][]{2009ApJS..184..218L}. Here we concentrate on so far unexplored redshift spikes at $z>0.8$. Table \ref{table:redshiftspikes} lists 13 significant redshift spikes in the range $0.6<z<4$, which are also colored in red in Figure \ref{fig8}. In order to guarantee sufficiently low false positive likelihoods, we required a minimum of eight members per $\Delta$z=0.005 redshift bin, and compared the detected number of members with the number expected from a $\Delta$z=0.1 bin, reduced by the objects detected in the redshift spike. Since the redshift ranges covered by the narrow-band filters are typically smaller than or comparable to the reference redshift range of $\Delta$z=0.1, the narrow-band filters create an artificial bias in this analysis, which could be mis-interpreted as redshift spikes. We therefore excluded the redshift ranges selected by prominent emission lines in the narrow-band filters. For the two Subaru Suprime-Cam narrow band filters NB711 and NB816, respectively, these are the redshift ranges $0.895<z<0.914$ and $1.164<z<1.208$ for the [OII] line, as well as the ranges $4.809<z<4.901$ and $5.637<z<5.771$ for the \Ly \ line, respectively. Indeed, one redshift spike was detected at $<z>=1.176$, which we excluded here. Table \ref{table:redshiftspikes} gives the number of objects detected in each spike, the average redshift and its standard deviation, as well as the expected number of objects and the Poisson likelihood for a statistical chance occurrence of the number of members. No correction for the relative areas subtended by the redshift spike features on the sky were made at this point. The last entry in table \ref{table:redshiftspikes} is a comment about the spatial distribution of the feature.

% TABLE 3:  Redshift Spikes 
% \scriptsize
\begin{deluxetable*}{lccccll}
% \tablewidth{0pt}
\tablecaption{Spectroscopically identified redshift spikes of large-scale structure}
\tablehead{
\colhead{Spike} &
\colhead{\#Members} &
\colhead{<z>} &
\colhead{$\sigma_z$} &
\colhead{Expected} &
\colhead{$P_{Poisson}$} &
\colhead{Comment}}

\startdata
z0667 & 69 & 0.6670 & 0.0013 & 25.2 & $7.1\times10^{-11}$ & several filaments\\
z0732 & 98 & 0.7321 & 0.0014 & 28.2 & $6.4\times10^{-11}$ & protocluster, several filaments\\
z0837 & 88 & 0.8373 & 0.0015 & 37.5 & $4.7\times10^{-11}$ & two clumps, filaments\\
z0891 & 98 & 0.8910 & 0.0015 & 37.0 & $5.6\times10^{-11}$ & two clumps\\
z0925 & 79 & 0.9253 & 0.0013 & 35.2 & $1.5\times10^{-10}$ \\
z1001 & 46 & 1.0014 & 0.0015 & 16.1 & $9.1\times10^{-10}$ \\
z1257 & 53 & 1.1257 & 0.0014 & 19.6 & $3.5\times10^{-10}$ \\
z1408 & 31 & 1.4080 & 0.0015 & 14.7 & $1.4\times10^{-4}$ \\
z1458 & 35 & 1.4579 & 0.0014 & 14.5 & $3.6\times10^{-6}$ & one filament\\
z1518 & 20 & 1.5185 & 0.0013 &  5.9 & $4.0\times10^{-6}$ \\
z1559 & 19 & 1.5586 & 0.0016 &  6.0 & $1.8\times10^{-5}$ \\
z2091 & 10 & 2.0906 & 0.0013 &  2.3 & $1.4\times10^{-4}$ & Virgo ancestor\\
z3982 &  8 & 3.9816 & 0.0013 &  1.7 & $3.9\times10^{-4}$ \\
\enddata
\label{table:redshiftspikes}
\tablecomments{Redshift spike members (column 2) are measured in a redshift interval $\Delta$z=0.005 and the expected number of objects (column 5) is derived from a redshift interval of objects $\Delta$z=0.1 (excluding the redshift spike). No correction for the relative area subtended by the structure is made. $P_{Poisson}$ is the Poisson probability of obtaining the observed number of members given the expectation value. Only features with $P_{Poisson}\leq4\times10^{-4}$ ($\sim3\sigma$) have been retained. The last column gives some comments about the geometry of the structures.}
\end{deluxetable*}
% \normalsize

\vskip 0.1 truecm
Figure \ref{fig9} shows the sky distribution of the objects in three of the thoirteen redshifts spikes discussed above. In the case of the $z=1.458$ spike there is a strong concentration of $\sim 18$ objects in an elongated structure of $\sim 12 \times 4 \ {\rm arcmin}^2$ ($\sim 6 \times 2 \ {\rm Mpc}^2$). In our sample this is the best example of a protocluster, which may turn into a rich cluster in the future. The DM halo mass of the these structures can be estimated through mock catalogs from simulations \citep[e.g.][]{2013ApJ...765..109D}. This, however, is beyond the scope of this paper and will be done in a future publication. The middle panel of Figure \ref{fig9} shows a redshift spike of ten objects at $z=2.091$.  Six of these are X-ray sources, and four have been detected by Spitzer and Herschel in the mid-IR. Because of the difficult redshift range most of their redshift (8 out of 10) identification qualities are relatively low (Q=1.5). Nevertheless, this structure is spatially consistent (both in redshift and sky distribution) with belonging to     
the Virgo-like cluster ancestor at z = 2.095 discussed above. The centers of the subclumps identified by \citep{2014ApJ...795L..20Y} are indicated by blue triangles in the figure. The right panel shows the redshift spike of eight objects at $z=3.982$, the least significant and most dispersed of our candidates. Follow-up observations and comparisons with other spectroscopic catalogues are neccessary to confirm the nature of these concentrations.

\section{Spectroscopic Confusion and \Ly \ Lensing}
\label{secdual}

Spectrocopic blending of galaxies at different redshifts in the same slit is a key source of uncertainty in future weak lensing cosmology experiments such as Euclid, LSST, and WFIRST \citep[e.g.][]{2017arXiv171008489R,2016ApJ...816...11D,2015APh....63...81N}.  Specifically, blended objects skew the photometric redshift distribution of weak lensing tomographic bins, leading to biases in the mean redshift distribution that must be corrected at the 0.2\% level for stage IV dark energy experiments. Even if objects are at the same redshift, the blending leads to biases in shape measurement which need to be corrected.  Therefore, constraining the fraction of blends with different and similar redshifts is key for these experiments. Estimates of possible blends, based on photometric counts, range from 1\% in DEEP2 \citep{2015APh....63...81N} to as high as 50\% \citep{2014wps..prop....4S} at the depths of WFIRST and so distinguishing the number of blended objects which have measurably different redshifts is important. 

\vskip 0.1 truecm
In our spectroscopic sample we find 43 objects with evidence of two different redshifts in one spectrum. This is based on an interactive visual inspection of the spectra, and thus affected by biases. Typically, only strong emission line spectra can be identified as these kind of interlopers. In some of these cases we can clearly see the presence of two galaxies in the same slit from the HST ACS images. Compared to our sample of 5515 high-quality spectra of galaxies with magnitudes $19<I<25.3$ (the LSST lensing depth) this yields a fraction of $0.8\%$ observed spectroscopic interlopers. This should be viewed as a lower limit since there is no guarantee we would obtain a redshift for the second source if it does not have strong emission features in the spectral range covered. 

\vskip 0.1 truecm
Another constraint on the potential fraction of interlopers with different redshift can be obtained from the percentage of catastrophic outliers in the comparison between spectroscopic and photometric redshifts. For this comparison we use the 3700 galaxies with the highest spectral quality (Q=2) and magnitudes $19<I<25.3$, for which galaxy templates yield the best photometric redshift solution determined by \cite{2016ApJS..224...24L} or \cite{2009ApJ...690.1236I}. Because of the additional photometric redshift model selection, this is a subsample of the 5515 high-quality spectra in the same magnitude range discussed above. Of these, 198 or $5.4\%$ have catastrophic photometric redshift outliers $|\delta z/(1+z)|>0.1$. On one hand, this is likely a lower limit, because the two blended sources need to have roughly similar brightness and a large enough redshift difference, in order to yield a significant error on the photometric redshift. On the other hand, the fraction of photometric blends could be larger than that of spectroscopic blends, because the spectroscopic slit reduces the cross-section for a blending impact. Also, we compare here the fraction of real overlapping sources with the fraction of catastrophic errors in photoz. The latter are due to a mix of two different aspects - one is the projected overlap, that may confuse the photometry, the other is the inevitable degeneracy in the photo-z solution which may lead to catastrophic outliers. This degeneracy is dependent on the depth and number of filters available, so it is difficult to draw any general conclusion from these numbers. Both effects discussed above, however, give an upper limit constraint on the true fraction of photometric confusion. Interestingly, the fraction of catastrophic photometric redshift outliers among the 50 identified spectroscopic blends is about 11\%, about twice that in the general high quality sample. These fractions of spectroscopic blends can be compared to the expected rate of 14\% for all blends, including those at the same redshift \citep{2016ApJ...816...11D}.

\vskip 0.1 truecm
The combined redshift distribution of the spectroscopic blends contains a surprising number of high redshift sources. About 28\% of the objects (12 out of 43) have redshifts in the range $4<z<6$, all identified with \Ly \ emitters. We can compare this with the sample of all serendipitous sources detected in our target slits, which represents the best approximation of a fair blind spectroscopic survey of field galaxies. We have 682 serendipitous galaxies with high quality spectra in our sample. Of these, only 8 objects (1.2\%) are at redshifts $z>3.7$, again, all of them \Ly \ emitters. In order to understand this large excess of blended \Ly \ emitters, we can look at the relative surface density of these objects in the different samples. 

\begin{figure}
\begin{center}
\includegraphics[angle=0,scale=.4]{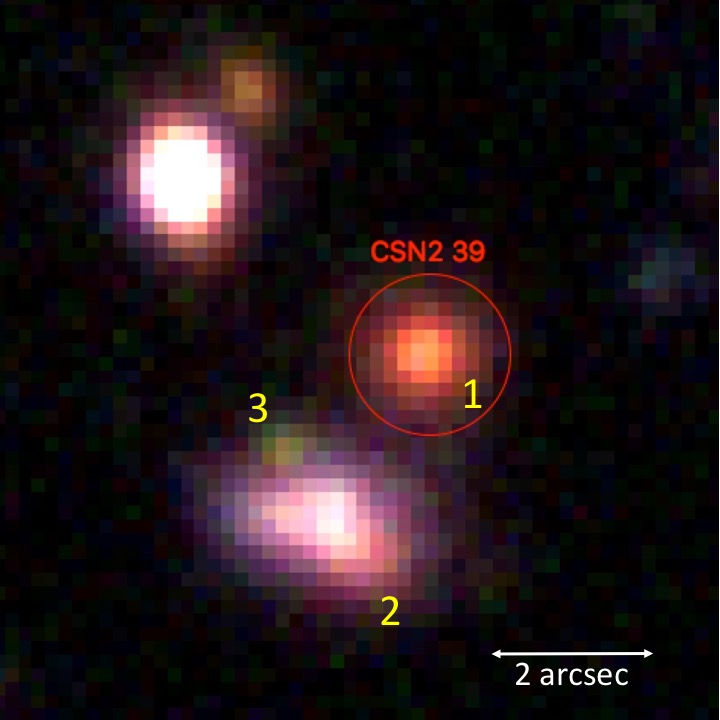}
\caption{Example of a $z=4.46$ \Ly \ emitter (object 3), putatively lensed by a $z=0.75$ galaxy (object 2), serendipitously detected in a slit targeted at object 1, a $z=1.22$ emission line galaxy. The angular separation of the lensed object from the lens is $\sim1\arcsec$. The RGB image is composed of the ultradeep Subaru HSC g- (blue), r- (green), and i-band (red) data in the COSMOS field \citep{2017arXiv170600566T}. The \Ly \ line is seen only in the r-band. \label{fig10}}
\end{center}
\end{figure}

\vskip 0.1 truecm
The surface density of serendipitous \Ly \ emitters can be estimated from the total blind spectroscopic area in our survey, which is the number of slits (about 7000 slits have been analyzed for serendipitous sources), multiplied by the average slit length ($\sim 12\arcsec$) times the average slit width ($\sim 1\arcsec$). This yields a blind survey area of $\sim 23 \ arcmin^2$, and thus a serendipitous \Ly \ source density of $\sim 1250 \ deg^{-2}$ over the redshift range $3.5<z<6$. In comparison, the LALA survey \citep{2007ApJ...671.1227D} has spectroscopically identified 73 \Ly \ emitters in the redshift range $4.37<z<4.57$ in a field of $0.7 \ deg^2$. In this particular redshift shell the surface density is thus $\sim 100 \ deg^{-2}$. Extrapolating this to the redshift range $3.7<z<6$ using the \Ly \ emitter luminosity function compiled by \cite{2014ApJ...788...87F} yields a surface density of \Ly \ emitters of $\sim 1000 \ deg^{-2}$, comparable to our blind spectroscopic survey. This is in stark contrast to the 12 \Ly \ emitters blended with other galaxies in our survey. Again, the survey area can be estimated by the number of slits searched (in this case corresponding to the total number of reasonable quality spectra in our sample, i.e. $\sim 7800$), multiplied by the average size of objects on the slits (estimated to be $\sim 4\arcsec$) times the slit width $1\arcsec$. This yields a survey area of $8.7 \ arcmin^2$, and thus a surface density of $\sim 5000 \ deg^{-2}$, about a factor of 4 higher than in the blind field survey.

\vskip 0.1 truecm
One possible interpretation of this excess is lensing and thus magnification bias of background \Ly \ emitters by the foreground galaxies targeted on the spectroscopic slits. One particular example, where we could localize the background \Ly \ emitter about $1\arcsec$ offset from the center of the putatively lensing foreground galaxy is shown in Figure \ref{fig10}. The median \Ly \ flux of our blended source sample is $\sim 10^{-17} \ erg \ cm^{-2} \ s^{-1}$. Around this flux, the slope of the cumulative number counts of $4<z<6$ \Ly \ emitters is estimated to roughly $\Gamma\sim -1.4$, again, extrapolating the luminosity function in \cite{2014ApJ...788...87F}. A factor of 4 increase of the surface density of \Ly \ emitters thus roughly corresponds to an effective flux limit a factor of $\sim 2.7$ lower. A significant part of this could indeed be due to strong lensing, if the lensed objects are situated well withing the Einstein radius of their lenses. The average strong lensing magnification over the Einstein radius is about a factor of 2. At about 40\% of the Einstein radius the magnification is 2.7. Most of the foreground galaxies of our blended \Ly \ emitters have stellar masses $8<log(M/M_\odot)<11$ and redshifts $0.7<z<1.5$. Together with the redshifts of the lensed objects in the range $4<z<6$, and assuming a mass to light ratio $\sim200$ and a fraction of 10\% for the dark matter mass within the projected Einstein radius, this yields Einstein radii in the range $0.1<R_E<2.3\arcsec$, with a median around $0.9\arcsec$. Since our slit width is typically $1\arcsec$, it is conceivable, that we detect strongly lensed objects well within the Einstein radii of their lenses. There are, however, aspects of caution: Some fraction of the putative \Ly \ blends could be artefacts, like e.g cosmic rays or sky subtraction issues. Also, assigning the interloper designation to the higher of the two blended redshifts could introduce some redshift bias.

\section{Summary}
We present a catalog of 10718 objects in the COSMOS field observed through multislit spectroscopy with DEIMOS on the Keck II telescope. The objects have been selected from a variety of input catalogs based on multi-wavelength observations in the field, and thus have a diverse selection function. We have a success rate of 62\% for high-quality spectra in the overall field and obtain a broad redshift distribution up to $z<6$ with peaks at $z\sim1$ and $z\sim4$.

\vskip 0.1 truecm
A direct object-to-object comparison with a multitude of other spectroscopic samples in the same field shows that our DEIMOS sample is among the best samples in terms of the fraction of discrepant spectroscopic redshifts ($\le 1.8\%$), and relative redshift accuracy of $<\delta z/(1+z)> \le 9\times10^{-4}$.

\vskip 0.1 truecm
We have identified 13 redshift spikes at $z>0.65$ with chance probabilities $<4\times10^{-4}$, some of which are clearly related to protocluster structures of sizes $>10\,{\rm Mpc}$. 

\vskip 0.1 truecm
We have determined the fraction of spectroscopic blends to about 0.8\% of our sample. This is likely a lower limit and at any rate well below the most pessimistic expectations. Interestingly, we find evidence for strong lensing of \Ly \ background emitters within the slits of 12 of our target galaxies, increasing their apparent surface density by about a factor of 4.

\acknowledgments
We would like to thank the anonymous referee for very useful feedback which helped to improve the presentation significantly.
Support for this work was provided in part by NASA through ADAP grant NNX16AF29G. AJB acknowledges support from NASA ADAP grant NNX14AJ66G and NSF grant AST-1715145. We would also like to recognize the contributions from all of the members of the COSMOS Team who helped in obtaining and reducing the large amount of multi-wavelength  data  that  are  now  publicly  available through the NASA Infrared Science Archive (IRSA) at http://irsa.ipac.caltech.edu/Missions/cosmos.html. This research has made use of the NASA/IPAC Extragalactic Database (NED) which is operated by the Jet Propulsion Laboratory, California Institute of Technology, under contract with the National Aeronautics and Space Administration. We thank the IfA graduate students Jason Chu and Travis Berger for observing some of our multislit masks with DEIMOS. The authors wish to recognize and acknowledge the very significant cultural role and reverence that the summit of Maunakea has always had within the indigenous Hawaiian community. We are extremely grateful to have the opportunity to conduct observations from this mountain.  
%\facility{facility ID}
\facilities{W. M. Keck Observatory; DEIMOS Spectrograph} 
\software{Dsimulator, DEEP2 pipeline, SpecPro}

\bibliographystyle{yahapj}
\bibliography{references}

\end{document}